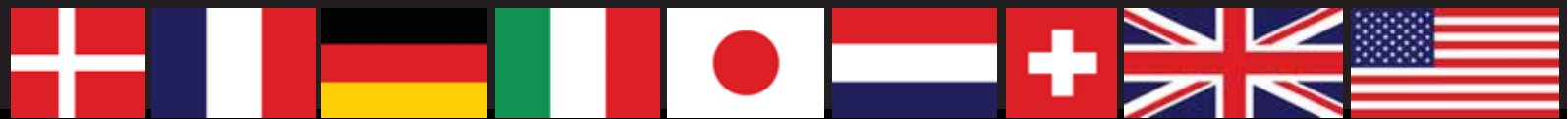

# EDGE

**E**xplorer of
**D**iffuse Emission and
**G**amma-ray Burst
**E**xplosions

*Luigi Piro*
*Jan-Willem den Herder*
*Takaya Ohashi*

## Participating Institutions:

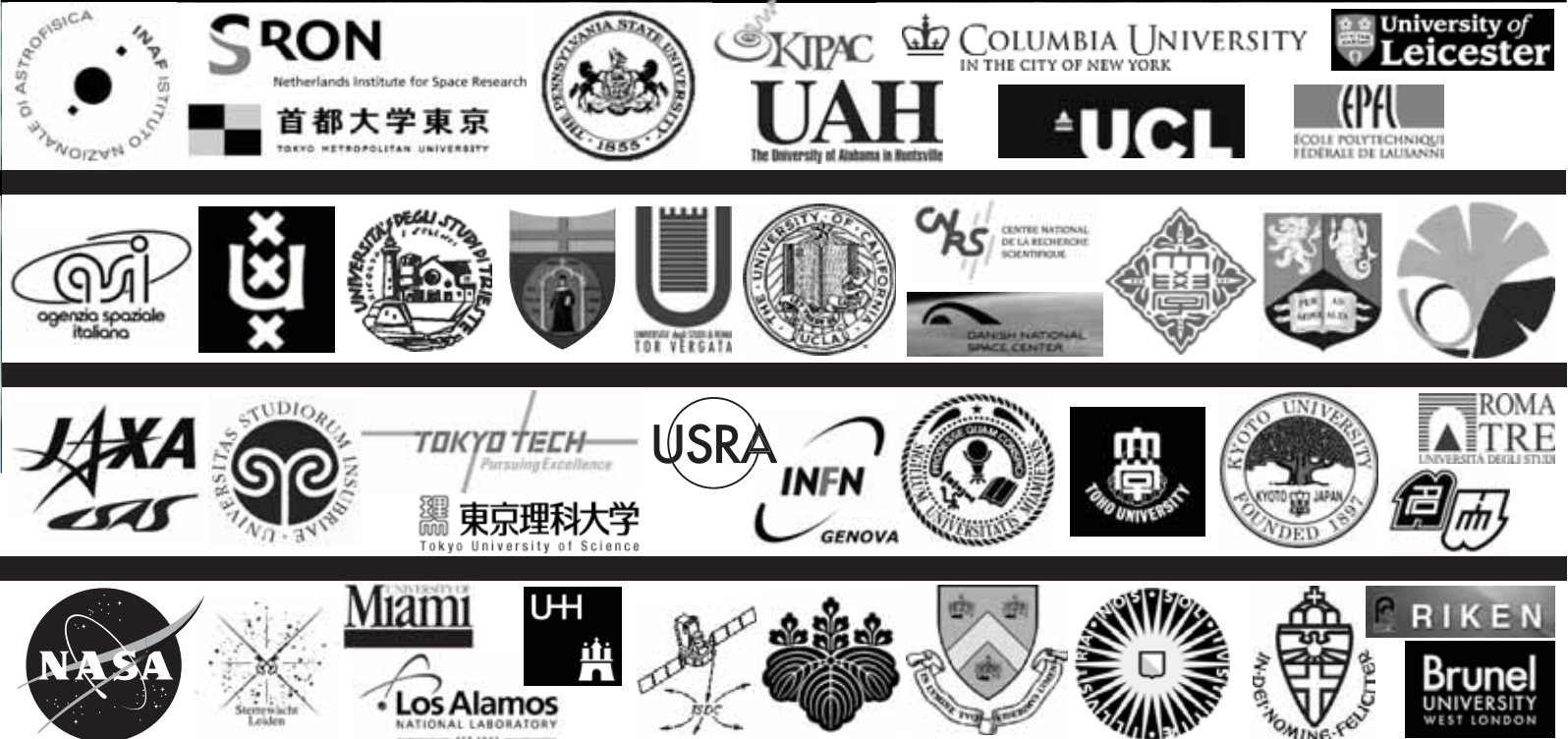

# EDGE: Formation and evolution of large scale structures in the Universe

Contents

---

This proposal has been prepared by a large group of scientists from the following institutes and in addition about 100 other scientists have shown their explicit interest in this mission.

Italy: INAF (Institutes and Observatories of Roma, Milano, Bologna, Palermo), Universities of Roma Tor Vergata, Roma 3, Bologna, Genova, Trieste, Palermo, Insubria (Como), Milano, INFN (Genova), ASI (ASDC)

Japan ISAS (JAXA), Tokyo Metropolitan University, Nagoya University, University of Tokyo, Tokyo Institute of Technology, Kanazawa University, Kyoto University, Tokyo University of Science, University of Tsukuba, Riken, Toho University

Netherlands: SRON, Universities of Amsterdam, Leiden, Nijmegen and Utrecht

Denmark: DNSC/Technical University of Denmark, Niels Bohr Institute/University of Copenhagen

Germany: University of Hamburg

France: CESR, Observatoire de Haute Provence, Saclay

Switzerland: ISDC, Paul Scherrer Institut, Institut des Hautes Études Scientifiques (ETH)

UK Brunel University, Birmingham University, University College London (MSSL), Leicester University

USA: MSFC, GSFC, University of Alabama in Huntsville, Penn State University, University Space Research Association, Los Alamos National Laboratory, Columbia University, University of Miami, KIPAC/Stanford, University of California at Los Angeles

**Science: WHIM:** E. Branchini, S. Borgani, A. De Rosa, M. Del Santo, Y. Ezoe, R. Fujimoto, M. Galeazzi, H. Kawahara, L. Moscardini, F. Paerels, M. Roncarelli, S. Sasaki, J. Schaye, Y. Suto, Y. Takei, E. Ursino, M. Viel, D. Watson, K. Yoshikawa. **Science: Clusters:** S. Molendi, S. Ettori, G. Brunetti, M. Girardi, L. Guzzo, A. Hornstrup, R. Lieu, P. Mazzotta, K. Pedersen, T. Ponman, P. Rosati, P. Tozzi. **Science: GRBs:** L. Amati , S. Campana, J. Atteia, S. Barthelmy, M. Boer, D. Burrows, A. Corsi, A. Galli, N. Gehrels, B. Gendre, G. Ghirlanda, D. Götz, J. Hjorth, N. Kawai, C. Kouveliotou, F. Nicastro, P. O' Brien, J. Osborne, G. Sato, R. Wijers. **Auxiliary science:** J. Kaastra, D. Barret, O. Boyarski, G. Branduardi-Raymont, M. Cocchi, A. Comastri, F. Haardt, J. In 't Zand, A. Kusenko, G. Matt, M. Méndez, M. Page, S. Paltani, O. Ruchayskiy, R. Salvaterra, R. Shaposhnikov, J. Schmitt, S. Sciortino, T. Tsuru, J. Vink, **other:** P. Giommi, N. White, S. Di Cosimo. **Instruments and Mission: WFS:** P. de Korte, K. Mitsuda, Y. Tawara, M. Barbera, L. Colasanti, G. Cusumano, L. Ferrari, F. Gatti, I. Hepburn, H. Hoevers, Y. Ishisaki, M. Macculi, T. Mineo, E. Perinati, A. Rasmussen, I. Sakurai, K. Shinozaki, N. Yamasaki, **WFI:** G. Pareschi, A. Holland, F. Christensen, D. Burrows, S. Basso, V. Cotroneo, P. Conconi, D. Spiga, G. Tagliaferri; **WFM+GRBDetector:** L. Natalucci, M. Briggs, S. Barthelmy, C Budtz-Jorgensen, E.Caroli, M. Feroci, M.H. Finger, J. Fishman, M. Kippen, C. Labanti, E. Quadrini, P. Ubertini.

We acknowledge the support of the **Italian Space Agency** and extensive and valuable help of **Thales Alenia Space** in the preparation of the technical part of this proposal. The proposal has been reviewed by an external committee with experts from Europe, Japan and the USA (G. Chincarini, W. Hermsen, K. Makishima, D. McCammon, S. Mitchell, R. Mushotzky, F. Paerels, C. Perola, T. Takahashi, H. Tsunemi, D. Willingale) and we wish to acknowledge their input.

# 1 Executive summary

One of the fundamental issues in astrophysical cosmology is to understand the formation and evolution of structures on various scales from the early Universe up to present time. EDGE will trace the cosmic history of the baryons, a key issue in the CV program (question 4.2), by measuring three tracers of cosmic structures:

Cosmic filaments

- Detect the largest reservoir of baryons from z~1 to the present time, predicted to reside in the Warm-Hot Intergalactic Medium (WHIM) by measuring densities down to $10^{-5}$ cm$^{-3}$ (~30 times smaller than currently probed within clusters of galaxies)
- Place constraints on the interplay between diffuse baryons and star formation

Clusters of galaxies

- Trace the evolution and physics of clusters out to their formation epoch (z>1)
- Measure the thermodynamical and chemical properties of a fair sample out to the virial radius, a fundamental step to qualify clusters as cosmological probes and for constraining their evolution through the link with the WHIM

Gamma-Ray Bursts

- Study the evolution of massive star formation using Gamma-Ray Bursts (GRBs) to trace their explosions back to the early epochs of the Universe (z > 6)
- Measure the metals in the host galaxies of GRBs and the explosive enrichment in their close environment out to z>6

This is illustrated in Fig. 1.1 where the unique capabilities of EDGE are shown.

In addition EDGE, with its unprecedented observational capabilities, will provide key results for a number of Cosmic Vision related science issues (questions 4.1, 4.3, 3.3, 3.2, 2.1) including: the study of feedback mechanisms into the Interstellar Medium (Supernova Remnants, galaxy/Active Galactic Nuclei outflows), constraints on the Dark Matter and Dark Energy content of the Universe (through clusters and GRBs), equation of state of the densest matter (neutron stars), GRB physics, upper limits on light dark matter particles, accurate measurement of the geometry of space-time by measuring X-ray afterglows of black hole mergers detected through gravitational waves, Active Galactic Nuclei and stellar population surveys, and Solar System physics.

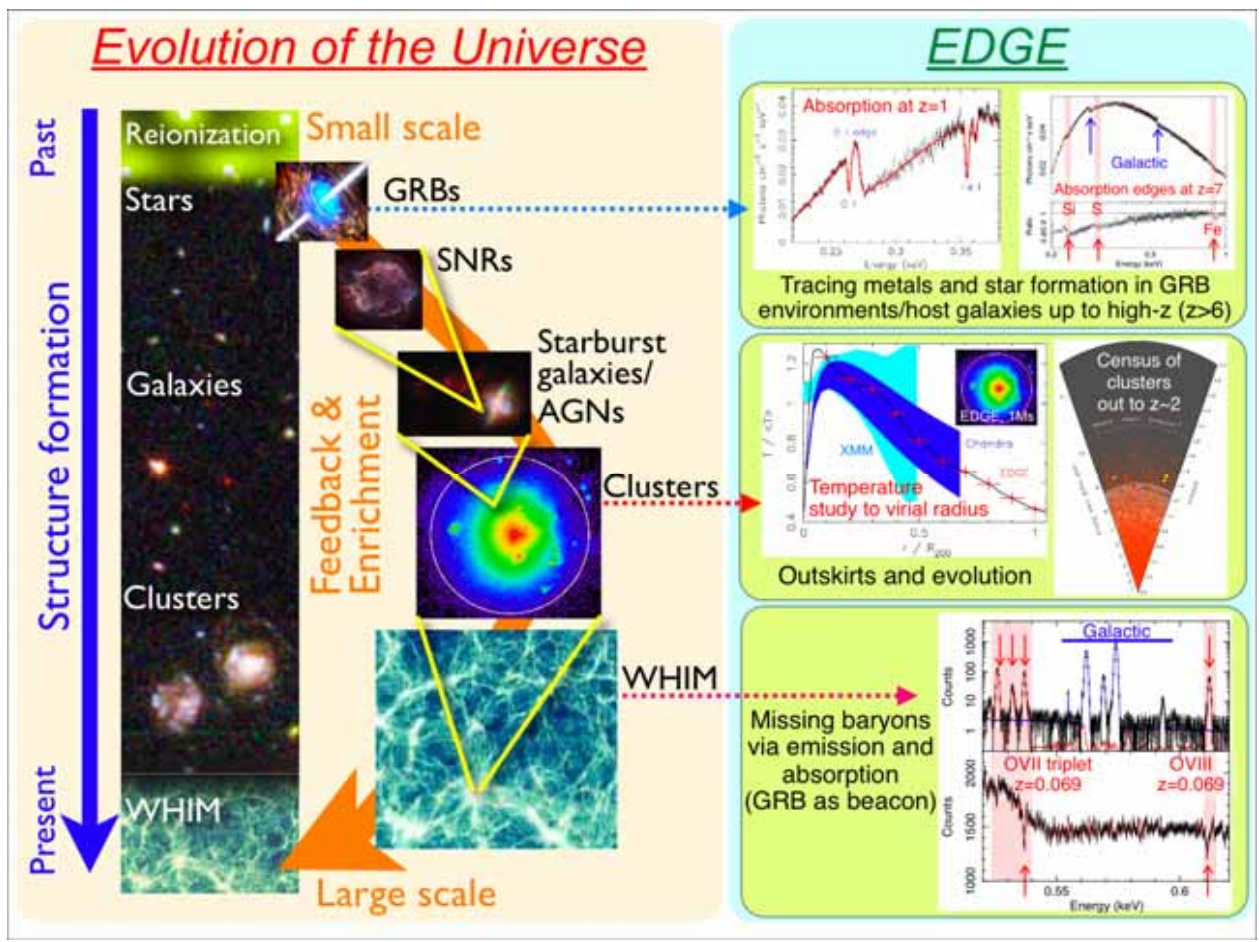

Fig. 1.1 History of structure formation in the Universe (left) and key EDGE capabilities (right)

## Science methods and uniqueness

Several outstanding contributions to the study of the cosmic history of baryons are unique to X-ray astronomy. EDGE is specifically designed to exploit the X-ray band-pass to investigate:

- Large scale, low density baryonic structures, including the WHIM and the outskirts of clusters of galaxies, which are visible only in X-rays. EDGE is uniquely positioned to observe them by high resolution spectroscopy and imaging. It will use GRBs as bright backlight beacons.
- The early populations of massive stars that ignited in the Universe and cannot be observed individually by any planned facility. EDGE will observe their explosive death and reconstruct the exact epoch of the first significant Fe enrichment, which is expected to signal the very first massive star explosions.

## Science instruments

The science is feasible using existing technology combined with innovative instrumental capabilities on a single satellite (see Fig. 1.2) comprising four instruments:

- Wide Field Spectrometer (WFS): effective area 1000 cm$^2$, 3 eV resolution, 0.7 x 0.7 deg$^2$ Field of View (FoV), optics with 2/4 reflections and TES calorimeter as detector
- Wide Field Imager (WFI): effective area 1000 cm$^2$, 15" angular resolution constant over the full 1.4$^o$ diameter FoV, CCD detectors
- Wide Field Monitor (WFM): FoV ¼ of the sky, consistent with requirement for follow up measurements of 80 bright bursts per year
- Gamma-Ray Burst Detector (GRBD): extension of energy range of the instruments to 3 MeV

With the combination of the WFImager and WFSpectrometer we will perform high resolution spectroscopy with the lowest possible background. The WFMonitor and GRBDetector will trigger the fast repointing of the satellite within one minute.

## Observations and Data policy

The driving science calls for a significant **core program** (80% in the first 3 years) to detect and characterize the Warm-Hot Intergalactic Medium, clusters of galaxies and Gamma-Ray Bursts. To reach the required sensitivity for these low surface brightness objects we typically need 1 Ms observations or fast follow up of transients. The auxiliary science will benefit from the core program data. In addition we plan a **guest observer program** (20% in first 3 years, later more) to exploit the unique capability of EDGE. Except for the guest observer program, where the standard 1 year proprietary data policy is followed, the data of the core program will be open to the community.

## Mission profile

EDGE is feasible as a truly ESA mission within the cost cap of an ESA medium class mission. Its fast repointing (1 deg/s) requires a compact satellite. With a densely packed payload this is compatible with the VEGA launcher. Its low background requires a low Earth equatorial orbit. We have adopted a lifetime of 3 years, appropriate to realize the major goals of the mission with a modest guest observer program. An extension to 5 years is feasible within the budget constraints.

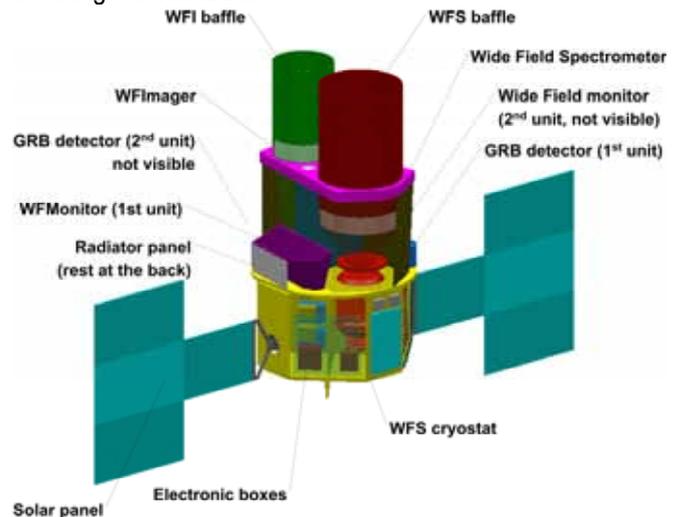

*Fig. 1.2 EDGE satellite and payload*

## Timeliness of EDGE

EDGE is highly complementary to operational and planned missions:

- the only mission with high spectral resolution (< 3 eV), large GRASP (> 400 cm$^2$deg$^2$) and fast repointing
- 2 orders of magnitude improvement in line detection sensitivity for extended sources compared to existing and planned X-ray missions (Fig. 3.2)
- point source sensitivity better than the XMM/Lockman Hole observation for a sky coverage of >10 deg$^2$ (XMM/Lockman Hole: 0.25 deg$^2$) (Fig. 3.3)
- 1 order of magnitude lower instrumental background than current missions and future observatory X-ray missions (XEUS, Con-X)
- vast increase in redshift coverage through high resolution spectroscopy up to the formation epoch for clusters of galaxies and earliest star formation
- first time high spectral resolution for transient sources (minutes)

The technology is well developed in Europe (Technical Readiness Level ≥ 4) and little risk is involved.

## The EDGE collaboration

The EDGE collaboration includes national agencies and major institutions from Italy, Japan, Netherlands, Denmark, France, Germany, Switzerland, UK and the USA. The payload will be provided by institutes from ESA member states, Japan and the USA.

# 2 Science

## 2.1 Introduction

Most of our knowledge about the formation of cosmic structures on various scales is based on observations of baryons locked in relatively dense gas accumulated in stars, galaxies, and clusters of galaxies. Surprisingly, current state-of-the-art observations only account for about 30% to 50% of the baryons at redshift z~0, while the rest of them is thought to be in diffuse, highly ionized, intergalactic gas that permeates the cosmic web, the so-called Warm-Hot Intergalactic Medium. Therefore the WHIM retains key information on the history of gravitational collapse and heating at the accretion shocks, and on the time-dependent kinetic energy injection from galactic winds and AGN jets. The elemental abundances of this gas reflect the history of metal enrichment from the very early times.

Thanks to its unique observational capabilities, EDGE will be able to study the role of the baryons in of the Universe from the early epochs, through Gamma-Ray Burst (GRB) explosions, via the period of cluster formation, down to very low redshifts.

The unique way in which EDGE will contribute to this study is schematically illustrated in Fig. 1.1. Due to the formation of the first stars, protogalaxies, and massive black holes the Universe reionizes. As cosmic structures take shape, gravitational collapse of Dark Matter halos creates the seed regions where star formation occurs; subsequently galaxies form and merge into groups and clusters through a hierarchical build-up process.

EDGE traces this evolution by observing the death of massive stars that explode as GRBs back to the first generations in the dark Universe. The associated supernovae inject nucleosynthesis products and energy into their environment. EDGE will detect this first metal enrichment through X-ray absorption spectroscopy of bright GRB afterglows.

Clusters of galaxies are the last and most massive structures to collapse. Besides containing a significant portion of the total mass, the outskirt regions of clusters show the most visible signs of the hierarchical formation of cosmic structures, of the accretion shocks and turbulence injection, as well as of the metal enrichment from relatively recent star formation. EDGE will conduct high sensitivity imaging and high-resolution spectroscopy out to the virial radius, thus performing a detailed thermo- and chemo-dynamical study of diffuse baryons inside and around clusters. Moreover, EDGE will carry out surveys that will detect the formation of the first cluster-sized objects thereby placing tight constraints on the values of cosmological parameters by tracing the growth-rate of cosmic density fluctuations.

As we approach the current epoch, between a third and one half of the baryons are expected to reside in the WHIM, which is currently undergoing gravitational collapse. The bulk of this medium remains still undetected. EDGE will observe the WHIM and measure its properties via absorption and emission in transitions of highly ionized metals. These observations close the baryon budget at z=0 by constraining their density. In addition, EDGE will shed light on the star formation history from the imprint by the energy and metal injection left on the WHIM.

Accomplishing a survey and characterization of these components of the baryonic Universe will require high resolution soft X-ray spectroscopy and imaging over a wide field of view, with extremely low background and the ability to rapidly point at bright GRB afterglows. Currently no other mission under study will address these issues directly. The instrumental requirements follow naturally from the following considerations. *High resolution spectroscopy of bright, distant continuum sources* in the soft X-ray band will reveal the metals in the WHIM. We will use bright Gamma-ray Burst afterglows as our 'backlight' sources, or beacons, since they are an unlimited 'renewable resource', and occur out to very high redshifts (z > 7). For bursts with an afterglow fluence in the 0.3-10 keV band of $F \geq 10^{-6}$ erg $cm^{-2}$, the requirement of 1000 counts per resolution element necessary to detect 0.1 eV equivalent width metal absorption lines in the WHIM, naturally implies an effective area of 1000 $cm^2$ and a spectral resolution $\leq 3$ eV. As GRB afterglows fade quickly, we need a rapid repointing capability, with the spectrometer pointing at the source within 60 seconds after the trigger. To obtain a representative sampling of the WHIM, we require > 100 lines of sight at this sensitivity. Complementary to the absorption spectroscopy, we will *image the WHIM and the outskirts of clusters in the emission lines* of H- and He-like C, O, and Ne, and the L shell lines of Fe. Here, the spectral resolution is set by the required contrast of the extraordinarily faint emission against the grey background (instrument background, unresolved extragalactic point sources, galactic foreground, emission and, in the case of clusters, thermal continuum emission). The expected characteristic emission line intensity of about 0.05 photon $cm^{-2}$ $s^{-1}$ $sr^{-1}$ in the strongest O K shell line (out to redshift 0.3) leads to a required grasp for the *high resolution imaging spectrometer (Wide-Field Spectrometer)* of 400 $cm^2$ $deg^2$ and a spectral resolution of 3 eV. An angular resolution of ~4' matches the typical size of WHIM filaments (1 Mpc~8' @ z=0.1). A low background is crucial for these measurements. This is achieved by the selected orbit, the low focal ratio for the telescope and the optimized detector shielding. The fraction of the cosmic X-ray background which is due to point sources (AGNs) can be reduced by a further factor ~3 if these point sources are known or if these can be identified and the relevant pixels of the imaging spectrometer can be rejected. This is achieved by a *high resolution imaging camera (Wide Field Imager)* with high contrast, large field-of-view and modest energy resolution. This instrument is co-aligned with the high resolution imaging spectrometer but has a good (HPD = 15") and constant point spread function over the field-of-view. This requires a detector with typical pixel sizes of 60 μm. This high angular resolution camera is very well suited to study the thermodynamical and chemical properties of clusters of galaxies out to the virial radius. The study of a representative sample of clusters, avoiding multiple pointings, requires typically a FoV of 1º, of the same magnitude as required for the high spectral resolution instrument. With an instrumental background as low as <1.5 $10^{-5}$ counts s$^{-1}$ arcmin$^{-2}$ due to the selected orbit and low focal ratio optics, a point source sensitivity of 1.5 $10^{-16}$ erg cm$^{-2}$ s$^{-1}$ for 0.5 – 2 keV will be achieved for a GRASP of 700 cm$^2$ deg$^2$ at 1 keV and 1 Ms observations. This improves the sensitivity by about a factor 100 with respect to previous missions. The 15" PSF matches the confusion limit at this sensitivity (1/25 sources/beam). The truly diffuse X-ray background will be characterized by using the data from the imaging camera (WFI), from the high resolution spectrometer (WFS) in combination with additional off-source measurements with a similar sensitivity.

Using GRBs to reveal metals in the WHIM and to trace the first stellar evolution from the local Universe up to high redshift (z>6) requires *rapid localization and re-pointing*. To carry out follow-up observations with high resolution spectroscopy of ~150 bursts over the mission lifetime with sufficient afterglow fluence (> $10^{-6}$ erg cm$^{-2}$ in 0.3 – 10 keV) we need a *Wide Field Monitor* (WFM) with a field-of-view of 2.5 – 3 sr and a sensitivity of 0.5 photon cm$^{-2}$ s$^{-1}$ (1 s, 5σ, 15-150 keV). With this sensitivity we expect to include in our sample of GRBs with bright afterglow ~7 events beyond z>6. By including a *detector with high energy sensitivity* (Gamma-Ray Burst Detector) (>1 MeV), we can measure the broad-band energy output of the GRBs. If the empirical correlation between the spectral peak frequency and isotropic luminosity, combined with the measured redshift, is confirmed using a larger sample than currently available we can construct the Hubble Diagram of GRBs. This provides geometrical constraints on cosmological parameters.

Along the way, we can study the astrophysical processes associated with the injection of energy and metals ('feedback') in great physical detail: we will obtain high resolution imaging and spectroscopy of supernova remnants, learn about the element enrichment, as well as the physical properties of collisionless astrophysical shocks similar to those responsible for heating up the WHIM. We will obtain emission spectra of starburst galaxies and AGNs, and measure their net kinetic energy luminosity as well as their mass loss.

The unique capability of EDGE is visualized in Fig. 2.1.1 using the phase diagram of the baryons at z=0 as predicted by a cosmological hydrodynamical simulation. For gas in systems ranging from the lowest density phase of the Intergalactic Medium (IGM), to the densest gas in star-forming galaxies, we plot normalized density (ρ/<ρ>=δ+1) and gas temperature. Current observations are effectively limited to the cooler parts of this diagram (T<$10^5$ K) and to the hotter and denser parts (clusters of galaxies T>$10^7$ K, δ>300). EDGE will push this into yet unexplored regions.

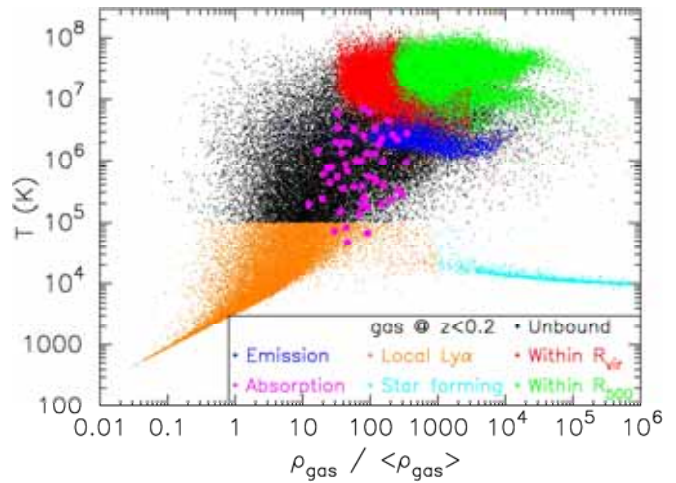

*Fig. 2.1.1 The EDGE contribution to the sampling of the phase diagram of cosmic baryons. EDGE will extend the current observations of the central regions of galaxy clusters (green) out to their outskirts (red) and to the dense part of the WHIM (blue), down to the regions with lower density (purple).*

In addition to the driving science, which we will describe in detail in Section 2.2, the mission will also provide a wealth of data for many other fields. Part of these data are directly related to the Cosmic Vision themes, others only indirectly. A significant part comes for free from observations for science drivers. We will summarize the most important parts of these additional objectives in a section on 'Auxiliary science' (Section 2.3).

## 2.2 Driving Science

In this Section we describe the main science objectives: studying the formation of structure of the Universe from early epochs up to present time. We start with the low density nearby Universe and proceed up to much denser conditions in the early Universe. This science drives the instrument requirements.

### 2.2.1 Warm-Hot Intergalactic Medium

A careful inventory of the baryons (Fukugita & Peebles 2004) shows that at high redshifts, we have accounted for all the normal matter in the Universe. The baryon density in the Ly-α forest at redshifts 2-3 is consistent with the density predicted from Big Bang Nucleosynthesis and the measured abundances of the light elements, and, independently, with the measured fluctuation properties of the cosmic microwave background. But at the current epoch, somewhere between one third and one half of the baryon density is not accounted for by the matter observed in stars, gas, dust bound to galaxies, and the hot diffuse medium in galaxy groups and clusters. Cosmological hydrodynamical simulations suggest that the budget of baryons that seem to have 'disappeared' between z=3 and z=0, is accounted for by a diffuse, highly ionized, warm-hot intergalactic medium, preferentially distributed in large scale filaments connecting clusters and groups in the nearby Universe.

This gas is extremely hard to detect: H and He are fully ionized, but the thermal continuum emission is much too faint to be detectable against overwhelming backgrounds. The only characteristic radiation from this medium will be in the discrete transitions of highly ionized C, N, O, Ne, and possibly Fe: to detect the lines associated to such transitions requires high resolution X-ray absorption and emission line spectroscopy and imaging. The predicted fluxes are well below the reach of current instrumentation; a single possible detection of intervening O VII/VIII absorption towards Mrk 421 with the Low Energy Transmission Grating Spectrometer on Chandra was not confirmed in a deeper spectrum with the Reflection Grating Spectrometer on *XMM-Newton* (Nicastro et al. 2005, Rasmussen et al. 2006).

While mere detection of X-ray absorption and emission from highly ionized C, N, O, and Fe will reveal the presence of the 'missing baryons' our goal is more ambitious; we plan not just to detect but also to *characterize the physical state of the WHIM: its temperature, density, location and metal content which trace the evolution of the medium and its interplay with the history of star formation.*

We base our modeling of the properties of the WHIM and the requirements for EDGE, on large scale Dark Matter + hydrodynamic simulations, with a parameterized treatment of the effects of stellar feedback from star- and galaxy formation, chief among which is of course the metal enrichment. We use the simulations of Borgani et al. (2004) and Viel et al. (2004) to develop various WHIM models. We focus on the one which includes the large scatter in metallicity that characterize the model of Cen and Ostriker 2006. Their model successfully reproduces the observed properties of the intergalactic O VI absorbers (e.g. Danforth and Shull 2005), and is based on a phenomenological treatment of the physical distribution and state of the intergalactic metals, including the effects of galactic feedback and departures from thermal equilibrium. From the simulations we conclude that GRB after-glows, if repointed in one minute from their onset, will give a unique opportunity to measure the intergalactic absorption with high statistics and to constrain the full WHIM evolution, thanks to GRBs' distance.

Fig. 2.2.1 (bottom panel) illustrates a larger portion of the sample GRB afterglow absorption spectrum also shown in the left-bottom panel of Fig. 1.1 with intergalactic absorption lines marked. With 3 eV spectral resolution, 1000 cm² effective area, a wide field monitor with 2.5 sr FoV and a lower photon energy threshold E~8 keV to trigger on and localize X-ray Flashes we find, based on the *SWIFT* results, that there are ~30 bursts per year accessible to EDGE that will produce about 50 combined 5σ detections of O VII and O VIII resonance absorption lines at the same redshift (see Table 2.2.1). We have assumed a 60 s pointing delay, an equatorial Low Earth Orbit with appropriate Earth occultation statistics and South Atlantic Anomaly passages. With our chosen thresholds we will have less than one false detection over the expected 150 GRBs over the mission lifetime. We verified this by a Monte Carlo simulation. Of course, these numbers are critically dependent on theoretical uncertainties in the galactic feedback and metal enrichment mechanisms. Assuming a very conservative model which is still consistent with the O VI data we would still detect 12 systems per year. Should we realize our goal of 1 eV energy resolution, this estimate increases to 100 absorption line systems per year.

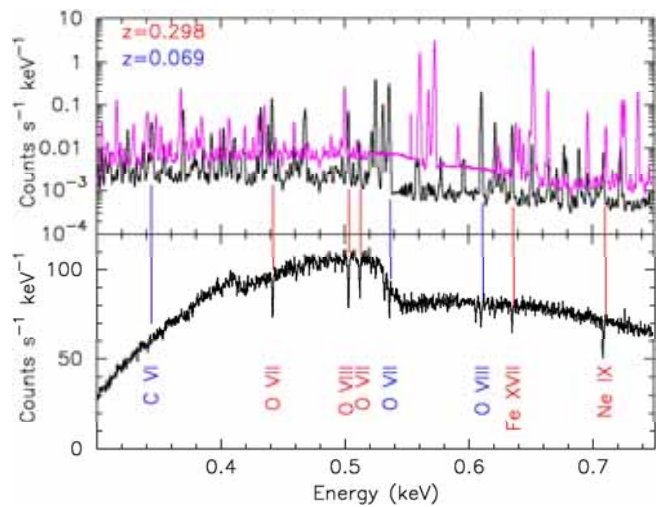

Fig 2.2.1 Emission spectrum of a 4 arcmin² area (top) and absorption (bottom) spectrum of the same region of the sky as measured by EDGE. In the top panel the emission of two red-shifted components is shown in black, whereas the emission of the Galactic foreground is shown in purple. In the bottom panel the spectrum of the same systems is shown but now in absorption using a bright GRB as a beacon.

*Table 2.2.1 Estimated number of absorption systems detected at >5σ significance per year by EDGE.*

| Fluence 0.3-10 keV [erg cm$^{-2}$] | # GRBs [yr$^{-1}$] | EW$_{min}$ O VII [eV] | EW$_{min}$ O VIII [eV] | # O VII/VIII [yr$^{-1}$] |
|---|---|---|---|---|
| >1 10$^{-5}$ | 6 | 0.11 | 0.09 | 22 |
| >5 10$^{-6}$ | 13 | 0.16 | 0.12 | 30 |
| >2 10$^{-6}$ | 33 | 0.25 | 0.20 | 50 |

From the absorption line spectroscopy, we will extract the O VII and O VIII column density distributions for systems with detections in both O VII and VIII, we sample the majority of the O atoms in the physical absorber.

The column density distribution determined from the absorption measurements, by itself provides a constraint on the models for the WHIM, but we propose in addition to detect, image, and measure the line emission. Simultaneous absorption and emission spectroscopy will allow for direct, model independent measurement of the characteristic density, ρ, length scale, ionization balance, excitation mechanism (or gas temperature), and element abundance of an intergalactic gas filament. In addition, the imaging emission line spectroscopy will provide unique 3D spatial information, while the absorption line spectroscopy will be sensitive to the lowest density absorbers (emission line strength scales as ρ$^2$, while absorption line equivalent width scales as ρ).

For the emission line simulation, we assume a spectral resolution of 3 eV, which is dictated by the requirement of continuum background suppression and the ability to distinguish the Galactic foreground thermal line emission. With our small focal ratio, we can match the physical size of the TES detector elements to a meaningful angular detection element on the sky. With a 1000 cm$^2$ effective area and a 120 cm focal length telescope, we will have 4 arcmin$^2$ resolution on the sky, well matched to the predicted characteristic angular scale of the WHIM. At 3 eV spectral resolution, we also resolve the He-like OVII n=1-2 'triplet', which makes the line identification very reliable and constrains the thermal emission rate and optical depth of the ion. With one more independent emission or absorption line measurement (O VIII, or O VII absorption against an afterglow), we have sufficient information to uniquely determine the complete physical state of the gas including metal abundance.

In a 1 Ms observation of a 0.7 x 0.7 deg$^2$ FOV, we predict detection of O VII emission line systems in about 1000 pixels. Fig. 2.2.1 (top panel) illustrates a sample emission spectrum of a 4 arcmin$^2$ area. The corresponding OVII emission line image is shown in figure 2.2.2. A mapping observation of a 2 – 3 deg field, encompassing the typical scale of the filaments, will effectively probe the 3D large-scale structure of the WHIM, including the effect of cosmic variance.

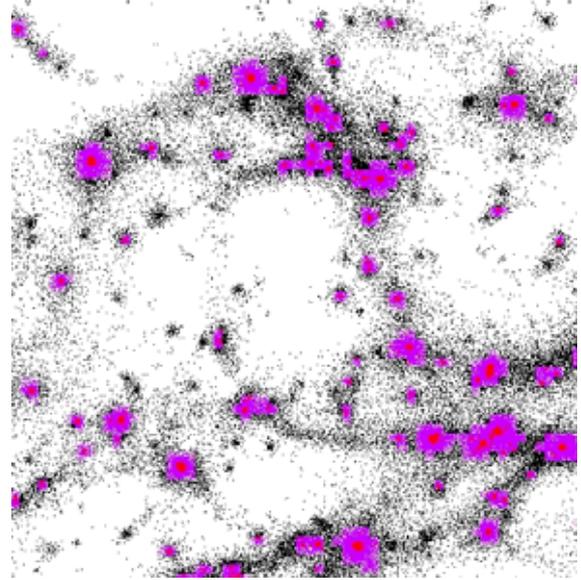

*Fig 2.2.2 Sky distribution of O VII emission relative to that of the gas, showing the simulated gas particles (black) and 5 σ detections of gas with overdensity δ> 1000 (red), and with δ < 1000 (purple). Area shown is 2.1 x 2.1 deg$^2$ centered on z=0.2 and for a 1 Ms observation.*

We expect to detect 18-25 % of the baryons down to overdensities of about 100 (Fig. 2.2.3).

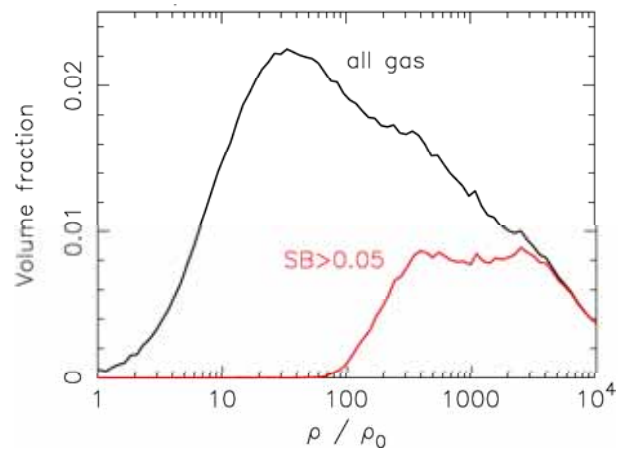

*Fig. 2.2.3 Volume fraction of all gas particles in the simulation of Fig. 2.2.2 (black) and for gas particles responsible for O VII emission lines with a surface brightness >0.05 photon cm$^{-2}$ s$^{-1}$ sr$^{-1}$ required for >5σ detections (red).*

At least one-third of the systems detected via absorption will be bright enough to be detected in emission and will allow us to perform a complete characterization of the physical state of the gas. A typical case is illustrated in the left-bottom panel of Fig. 1.1 in which the O VII and O VIII lines of the same system are seen in emission and absorption. Therefore, as part of our observing strategy, we plan to dedicate observations to fields centered on lines of sight for which we have obtained a high quality intergalactic absorption line spectrum. The most important source of confusion in the emission line data is the presence of misidentified fore- and background emission lines, and faint continuum point sources. The former will be suppressed by combination of two or more emission lines (O VII and O VIII) while the latter will be suppressed by correlation with the WFImager data.

### 2.2.2 Clusters of galaxies

Clusters of galaxies are the largest structures to have decoupled from the Hubble flow. While still carrying the imprints of primordial cosmological fluctuations they are undergoing processes where prodigious amounts of energy, second only to those associated to the Big Bang, are being converted from one form to another. While the bulk of the mass in clusters is in the form of dark matter, the dominating baryonic component is a hot tenuous X-ray emitting Intra-Cluster Medium (ICM). All observations of this medium to date have literally only sampled the tip of the iceberg: only regions within $R_{500}$, that is ~1.5 Mpc, or about 50% of the virial radius ($R_{200}$) for a $10^{15}$ $M_\odot$ object, have been well measured ($^4/_3 \pi \rho R_{200}^3 \equiv M_{200}$, $\rho=200\rho_c$, with $\rho_c$ the critical density of the Universe). Thus only a small fraction of the total volume of clusters has been mapped, missing those regions where a sizeable fraction, and for less massive systems in fact most, of the baryons reside and accrete from the field, and where the ICM is heated and enriched by metals. These low surface brightness regions are of course also organically connected to the WHIM.

Large uncertainties still affect the present reconstructed temperature profiles even at $R_{500}$ (see Fig. 2.2.5) and there are very few constraints available beyond that radius to precisely measure the mass within the virial region. Only sufficiently small errors allow the use of galaxy clusters as cosmological probes. A spatially resolved spectral analysis of the outskirts in nearby X-ray luminous objects is needed to recover the total physical properties of the ICM, and thus the fundamental properties of clusters, out to the virial radius. Furthermore, an estimate of the metallicity profile beyond $R_{500}$ will allow the evaluation of the total metal mass present in clusters, providing invaluable information on the cosmological star formation history, and on the corresponding initial mass function. While the observational limitations described above apply to massive and intermediate mass systems, the situation is even worse for poor clusters and groups, which are predicted to carry a substantial fraction of the cosmic baryon budget. For these objects, we currently only have a highly biased, incomplete census in the very local Universe and very little information on the outer regions.

We will measure *the surface brightness, temperature and metal abundance out to the virial radius* for an adequate sample of clusters, with medium spectral resolution imaging, providing temperature measurements from the thermal continuum and iron abundance measurements from the Fe L-shell blend. With a high spectral resolution, lower angular resolution imager, we can measure detailed emission line spectra out to almost as far, thus providing constraints on bulk motion and turbulence, and accurate abundances for the low-Z elements, mostly O, as well as for Fe.

These capabilities are illustrated in Fig. 2.2.4, where we show the vastly increased contrast for low surface brightness in the WFI compared to what is currently achievable. The high contrast will allow us to measure the cluster temperature out to the virial radius, see Fig. 2.2.5.

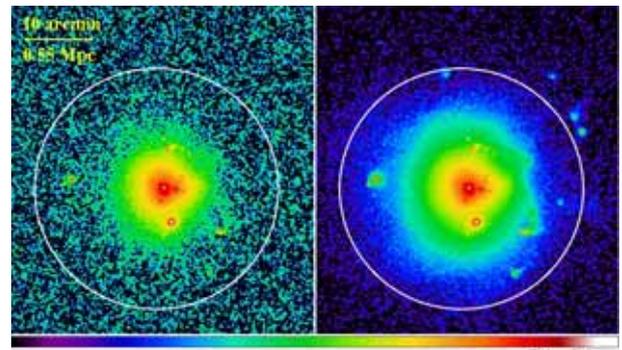

*Fig. 2.2.4 The vast improvement between XMM-Newton (left) and EDGE (right), mainly due to the better background and long observation times. The circle gives $R_{200}$.*

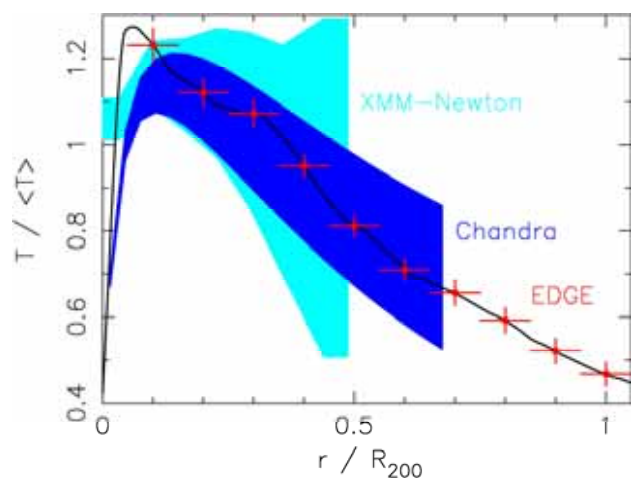

*Fig. 2.2.5 Typical ICM temperature profiles for current missions and expected EDGE measurements. The model (solid line) is from a hydrodynamical calculation of a massive system ($M_{virial}$= 2.1 $10^{15}$ $M_\odot$, Roncarelli et al., 2006). For this model $R_{500}$~0.65 $R_{200}$.*

The high resolution spectra give us unprecedented sensitivity to the physical conditions in the clusters, including the outskirts, where we will constrain the large-scale velocity fields associated with the past merger history of the cluster. We will see the hierarchical assembly of cosmic structures in action. High sensitivity to low surface brightness will characterize the regions upstream from merger shocks, while the high spectral resolution images will open the possibility of directly measuring gas velocity fields. By quantifying the amount of kinetic energy associated with bulk and turbulent motions, we will estimate the corrections to the equation of hydrostatic equilibrium, thereby providing the crucial calibration of mass measurement required for the use of clusters as tools for precision cosmology.

We plan to observe ~10 nearby luminous objects, both groups and clusters and with different known dynamical states mapping their emission up to the virial radius with a single exposure. Off-axis observations of the X-ray background will be used to provide an adequate characterization of its spectrum (Fig. 2.2.6).

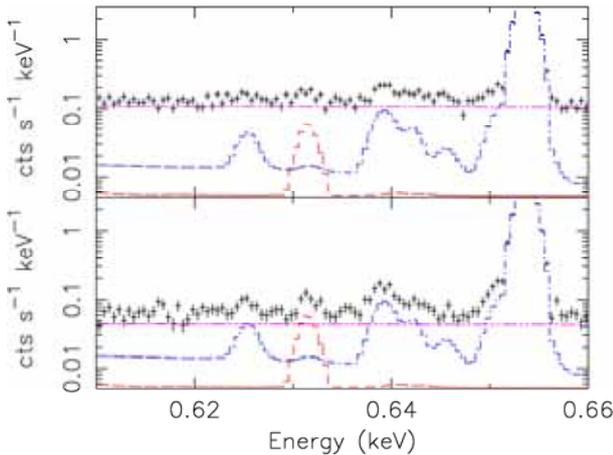

Fig. 2.2.6 WFS spectrum of a cluster outskirt (100 arcmin$^2$, 1 Ms, O abundance 0.15 times Solar) before (top) and after (bottom) excision of point sources identified in the WFI. The Galactic foreground is rendered in blue and the flat extragalactic background in purple. The significance of the redshifted OVIII line around 0.63 keV (red) improves after rejecting the point source contribution.

In addition to sampling the lowest surface brightness regions of nearby objects, EDGE will detect clusters to very high redshifts, as far back as their probable formation time. Chandra and XMM-Newton have provided gas distributions for a few massive systems which are already in place by z≈1. This suggests a formation age that dates back to z>1.5 (e.g. Rosati, Borgani & Norman 2002). We will address such fundamental questions as: (a) when and how do clusters form? (b) what is the thermodynamic history of the ICM, and what is the role of non-gravitational heating mechanism? (c) how is the ICM polluted by metals? (d) how do the cluster scaling relations take shape? To investigate these issues, EDGE will obtain a sample of clusters extending out to z>1. This sample will complement the results of future deep surveys in the optical/near-IR and millimeter bands (Sunyaev-Zeldovich effect), sampling large volumes at z>1.

To cover these science goals *we will perform three surveys at different depths* (see Table 2.2.2): a *Wide Survey* comprising a contiguous area of 100 deg$^2$, and ~240 non-contiguous pointings on GRB fields. Two deep surveys: *Deep-1* covers 12.5 deg$^2$, combining 8.5 deg$^2$ from fields used for the background controls for deep observations of individual clusters, and 3 deg$^2$ from four deep GRB follow-up pointings. *Deep-2* is the same dataset used to image the WHIM and covers a contiguous field of 8 deg$^2$.

Table 2.2.2 Sensitivities of deep surveys (3$\sigma$ flux limits $F_{lim}$ in $10^{-16}$ erg cm$^{-2}$ s$^{-1}$ for 0.5 – 2 keV).

|  | Wide | Deep-1 | Deep-2 |
|---|---|---|---|
| Exposure | 50 ks | 1 Ms | 2 Ms |
| $F_{lim}$ AGN Survey | 19 | 1.5 | 0.9* |
| $F_{lim}$ clusters | 30 | 4.5 | 2.7 |
| $F_{lim}$ with $T_X$ measure | 80 | 15 | 9.0 |
| Contiguous area (deg$^2$) | 100 | n/a | 8 |
| Total area (deg$^2$) | 340 | 11.5 | 8 |
| Clusters detected @ z>1 | 1800 | 510 | 600 |
| Clusters with $T_X$ @ z>1 | 450 | 140 | 170 |

* This is below the confusion limit of 1.5 $10^{-16}$ erg cm$^{-2}$ s$^{-1}$

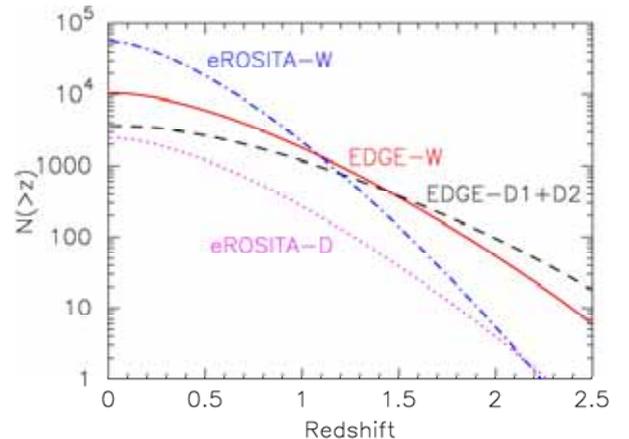

Fig. 2.2.7 Cumulative redshift distribution of EDGE surveys compared with the predictions for eROSITA. These predictions are based on the halo mass function computed for the best-fit WMAP-3yr cosmology and by extrapolating the evolution of the relation between mass and observed X-ray luminosity at z <0.6 (e.g. Vikhlinin et al. 2003; Lumb et al. 2004).

These surveys will detect about 3000 clusters at z>1, and measure the temperature for a subsample of about 800. We anticipate that EDGE will detect about 200 objects at z>2, a yet unexplored territory for the study of hot baryons. The cluster sample will also provide an estimate of cosmological parameters (Section 2.3.1). In Fig. 2.2.7 and 2.2.8 we give the cumulative redshift distributions for all clusters and for clusters for which EDGE will provide the X-ray temperature.

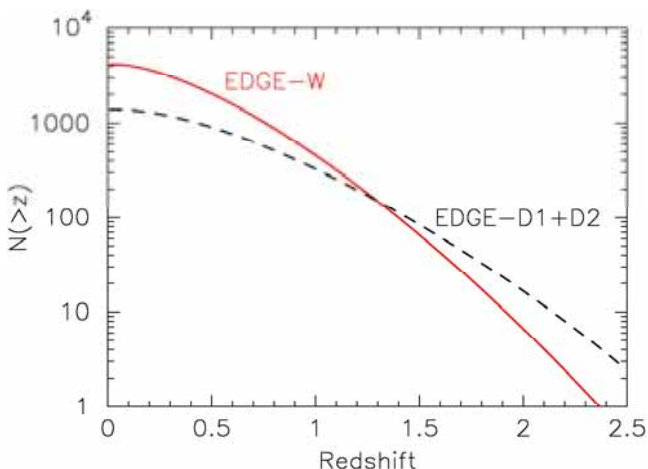

*Fig 2.2.8 Cumulative redshift distribution for EDGE surveys where the X-ray temperature is measured with at the most 30 percent error at a 90 percent confidence level.*

### 2.2.3 From the local to the Dark Universe with GRBs

It is well established that most long-duration gamma-ray bursts (GRBs) are caused by the explosive deaths of massive stars and, due to their enormous brightness and distances (Fig. 2.2.9) they can be seen throughout the observable Universe. Being characterized by copious amounts of penetrating high energy photons, they can probe the regions of the Universe beyond z>6, which are not accessible in the optical band. Given these facts, it has long been realized that GRBs could be powerful probes of star-formation activity throughout the history of the Universe. Indeed GRBs could be ideal tracers of star formation due to their brightness in X-rays, which are not heavily affected by dust extinction. They originate from a stellar explosion and thus do not require a detectable host galaxy. In the collapsar model, GRBs are expected to be formed only by stars with metallicity <0.3 of the solar value (MacFadyen & Woosley 1999). The redshift distribution of GRBs should thus map directly the star formation history of the Universe at z > 6, back to the first era of population-II objects, and possibly the first population-III stars preceding the formation of galaxies (e.g. Bromm & Loeb 2006).

Furthermore, as GRBs occur in rather typical galaxies, this provides a method to identify and study the types of galaxies that are responsible for the bulk of the star formation, and hence for the bulk of the (re)-ionization of the universe, in contrast to present and planned facilities, which favor the brightest systems.

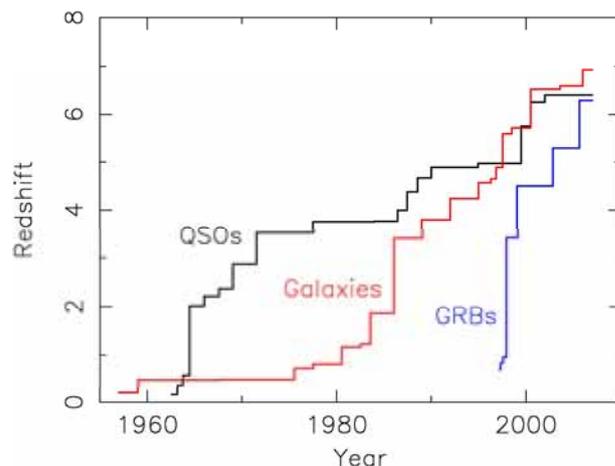

*Fig. 2.2.9 Redshift records for QSOs, galaxies and GRBs (adapted from Tanvir & Jakobsson 2006)*

The ~$10^6$ afterglow photons gathered by our instruments enable high resolution spectroscopy. One of our main goals is to use GRBs as cosmological beacons for the study of the star formation and of the metal enrichment history of the Universe out to z > 6.

This detailed study of GRB afterglows will open a completely new window on the formation and evolution of the baryons at high redshifts. While currently planned IR/sub-mm facilities (e.g. JWST, ALMA) probe the formation and evolution of the earliest massive protogalaxies, *EDGE directly reveals the processes associated with the first generation stellar evolution*, by detecting the events heralding the demise of the most massive first-generation stars. GRBs (in their long flavor) are indeed associated with SN-like explosions of massive stars and they show intrinsic (i.e. in situ) X-ray absorption typical of star forming regions. This absorption, which is primarily due to metals, indicates that the column density of metals is $A \times N_H = 10^{22}$ cm$^{-2}$ (Campana et al. 2006), where A is the metallicity relative to solar. It is indeed tantalizing that significant absorption is detected even in the highest redshift burst (z=6.3), indicating that metal enrichment was already vigorous at even earlier epochs. With high resolution spectroscopy, *EDGE will see directly the metal enrichment in the environment of the most massive stars*. At the highest redshifts we will see the medium-Z α-elements (Si and upward), and probe for the presence of Fe at these very early times. The exact epoch of first significant Fe enrichment is expected to signal the arrival of supernovae, and therefore probes the prior integral star formation rate. This is illustrated in Fig. 2.2.10 where we show the spectrum of a redshifted GRB (z = 7). Clearly the redshift and the column densities for the different edges can be accurately determined ($\Delta z = 0.02$ and relative error of column density 10-30%). Abundance patterns of metals whose edges are falling in the bandpass will be measured for bursts with afterglow fluence >$10^{-6}$ erg cm$^{-2}$ and for column densities $\geq 10^{22}$ cm$^{-2}$. At lower redshift we can also track the injection of the CNO elements, and derive absolute abundances (metallicity), because H and He dominate the opacity below the O edge. Measurements of metallicity can be extended to higher redshift by assuming the same relative abundance pattern of metals, i.e. CNO versus high Z elements, as measured at lower redshift by EDGE and by optical instruments (Savaglio 2006).

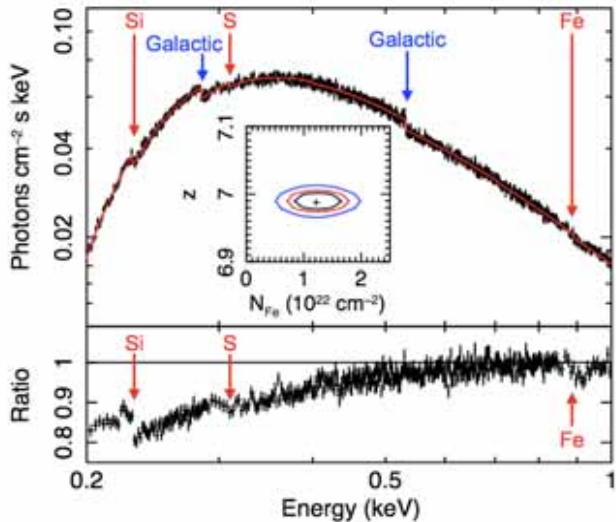

*Fig. 2.2.10 Spectrum of a z=7 GRB afterglow with a fluence of 4 $10^{-6}$ erg cm$^{-2}$ (0.3-10 keV), a column density of $N_H$ of 5 $10^{22}$ cm$^2$ and an abundance of 1/3 solar. Edges produced by Si, S, and Fe are clearly detected. Based on these edges the redshift and column densities can be accurately determined ($\Delta z \approx 0.02$).*

X-ray spectroscopy, contrary e.g. to the optical band, provides a direct measure of the column densities of elements in all physical and chemical states. The ISM in our own Galaxy, observed through bright X-ray binaries (Juett et al. 2006), has yielded dozens of detections of all abundant elements, in monatomic gaseous, molecular, and solid phases. The strongest absorption lines expected from the neutral or mildly ionized phase are 1s-2p transitions of C I-III, O I-III, Ne II-IV lines as well as K and L edges from the same elements. With EDGE it will be possible to apply this technique to characterize the ISM of GRB host galaxies (see Fig. 2.2.11). Expected EWs range from 0.1 to 1 eV for O I or Fe I for column densities down to $10^{21}$ cm$^{-2}$, so they can be detected for most of the bursts of our sample. The redshift of each line can be measured with an accuracy of $10^{-3}$, thus enabling kinematic studies of the gas.

The detection of resonant absorption lines from highly ionized metals, detectable only in soft X-rays, will enable the study of the warm-hot phase of the host galaxy Inter Stellar Medium (ISM) or host galaxy halo (already detected in our own Galaxy). They will be particularly important, in the context of large scale structure evolution, to disclose the presence of hot galactic outflows enriching the IGM well beyond the epoch when WHIM filaments formed. The most promising of such transitions, at high redshifts, are undoubtedly the $K_\alpha$ inner shell bound-bound transitions from mild- to high-ionization Fe (XVII and above).

A very high ionization phase is expected to develop resulting from the ionizing flux produced by the GRB in its close environment. This phase can be easily distinguished from the hot ISM in the host galaxy. Tracking the temporal evolution of absorption features (edges and lines) with EDGE will allow an unprecedented characterization of the circumburst medium (density, metal content and distance from the GRB), which in turn enlighten the origin of the progenitor.

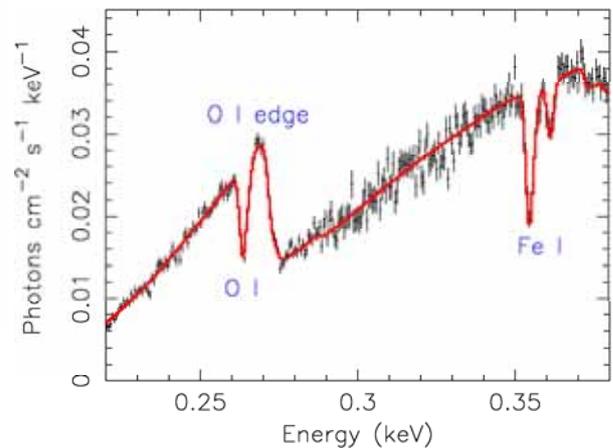

*Fig. 2.2.11 Spectrum of a GRB at z=1, $N_H$ (host galaxy) of 3 $10^{21}$ cm$^{-2}$, T=2 $10^4$K, 0.3-10 keV fluence 4 $10^{-6}$ erg cm$^{-2}$, integration time 50 ks.*

The instrument requirements are similar to those set by the requirements for detection of IGM by absorption spectroscopy (see section 2.2.1). *The goal is to gather a sample of about 150 GRB afterglows* (in 3 years) with 0.3-10 keV fluences >$10^{-6}$ erg cm$^{-2}$. This requires to localize and follow up about 250 GRBs. For this sample we will *measure the redshift, characterize the star formation, and study the history of metals in both the close GRB environment and its host galaxy up to high-z by X-ray spectroscopy.* Fast dissemination of the GRB coordinates, and particularly the early determination of the X-ray redshift measured by EDGE, will allow the community to carry out optical and IR follow up spectroscopy of the most interesting (i.e. high z) events. This will provide a sample of 'normal' high-z galaxies that will likely be missed in other surveys, but that once localized by EDGE, can be studied in detail by ALMA, JWST and ELT. Taking into account SWIFT results and theoretical predictions, we expect to localize and follow up about 20 burst at z>6 in 3 years, 7 of which will be characterized by an afterglow fluence >$10^{-6}$ erg cm$^{-2}$. This number can be doubled by lowering the nominal trigger threshold by about 50%.

## 2.3 Auxiliary science

### 2.3.1 Cosmological parameters

According to the emerging cosmological scenario, the Universe is dominated by non-baryonic Dark Matter (DM) and by an unknown form of Dark Energy (DE), whose relative amounts are such to cause an accelerated cosmic expansion at the present time, while making the Universe close to being spatially flat. The amount and nature of DM and DE affect both the global geometry of the Universe and the dynamics of structure formation. As we discuss in this Section, EDGE may provide geometrical constraints on the DE and DE density parameters $\Omega_M$ and $\Omega_\Lambda$, by investigating the use of GRBs as standard candles, and dynamical constraints, by means of the redshift distribution of the flux-limited samples of clusters of galaxies. Therefore, with a single experiment, EDGE will stringently constrain the "cosmic stress-energy tensor" by means of two independent and conceptually different methods.

Gamma-Ray Bursts: Because of their brightness and distance distribution (z up to 6.3, <z>≈2.6), GRBs are promising candidates to constrain cosmological parameters if, similarly to Supernovae (SN), they (or a subclass of them) can be proven to be standard candles. GRBs have a redshift distribution skewed towards higher redshifts than SN. Therefore at early epochs of the Universe when dark energy was supposedly starting to counter balance the gravitational pull of dark matter, GRBs provide information complementing that derived from SN only. This requires that the energy or the luminosity is precisely estimated from observable quantities. Correlations linking either energy or luminosity to observable quantities have been presented in the literature. Some of them are either too dispersed for precision cosmology (Amati et al. 2002, Amati 2006) or require long lasting follow-up campaigns to estimate the collimation corrected energy (Ghirlanda et al 2004). However a more promising correlation exists that links the properties of the prompt emission, namely the burst isotropic peak luminosity $L_{iso}$, the peak energy of the spectrum $E_p$ and a characteristic timescale $T_{0.45}$ (Firmani et al. 2006). Then only the redshift needs to be determined using the afterglow. However, there are a number of issues that have to be settled. The small number of bursts presently available prevents an assessment of possible systematics and selection effects. There is also a potential circularity problem when the same correlations, found by adopting a given cosmology, are used to constrain the cosmological parameters (Ghisellini et al. 2005). This can be cured with a relative large number of bursts (about a dozen) within a small redshift bin $\Delta z/z \sim 0.1$ (Ghirlanda et al. 2006).

Our goal is to collect a sample of 150 bursts with accurate measurements of redshift and of key parameters of the prompt emission in order to: a) study in detail the aforementioned correlations; b) if confirmed, apply them to the estimation of cosmological parameters. The precision of the cosmological constraints obtained through the above correlations depends on the number of events and on the accuracy of the observables. We have carried out an extensive set of simulations, adopting as reference the correlation linking $L_{iso}$ - $E_p$ - $T_{0.45}$ (including its present $1\sigma$ scatter of 0.06 dex). Experience with GRB experiments, and our simulations, show that a large energy band is crucial for an accurate estimate of $E_p$, requiring extension up to 1 MeV with an effective area of at least 500 cm² below 600 keV. The energy resolution provided by the EDGE instruments is more than adequate. This offers an accuracy of the order of a few % in the estimate of $E_p$ for typical bursts with 50-300 keV fluence >10⁻⁶ erg cm⁻² (corresponding to the ∼50% brightest GRBs). In order to study the impact of having a larger sample of bursts with precise measurements of their spectral properties we simulated 150 GRBs through the $E_p$-$L_{iso}$-$T_{0.45}$ correlation (see Ghirlanda et al. 2006 for the simulation details) with fluence >10⁻⁶ erg cm⁻² in the 50-300 keV energy range. This number corresponds to the observing goal of EDGE in 3 years. Fig. 2.2.12 shows the obtained 99% confidence level contours in the $\Omega_m$- $\Omega_\Lambda$ plane. As can be seen, with this method EDGE is expected to constrain these cosmological parameters with high precision.

Clusters of galaxies: In the standard cosmological scenario, clusters of galaxies form from the collapse of rare, high-density peaks of the primordial density fluctuations. For this reason their population and the distribution of their properties are highly sensitive to cosmological parameters. In particular, the mass function, i.e. the number density of clusters of mass M at redshift z, provides a potentially powerful tool to constrain cosmological models (Borgani & Guzzo 2001; Rosati et al. 2002; Voit 2005 for reviews). Clusters observed in X-rays offer a potentially unique way to trace this evolution: since the X-ray luminosity $L_X$ and temperature $T_X$ are related to the total collapsed mass M, the X-ray luminosity function (XLF) and the X-ray temperature function (XTF) can be used to infer the theoretically predicted mass function (e.g., Warren et al. 2006).

The advantage of using X-ray luminosity as a tracer of the mass is that $L_X$ is available by definition for all clusters observed in a flux-limited survey. However, since the X-ray emissivity depends on the square of the local gas density, $L_X$ is quite sensitive to any physical effect that affects the gas distribution. Indeed, analyses of ROSAT-based surveys have shown that systematic uncertainties, related to the uncertain relation between $L_X$ and M, are already three times larger than the statistical uncertainties for samples containing ∼100 clusters (Borgani et al. 2001; Vikhlinin et al. 2003, Schuecker et al.

2003). $T_X$ is in principle a closer proxy to the cluster mass, through the assumption of hydrostatic equilibrium. However, precise $T_X$ measurements are at present available only for a limited number of distant objects. Clearly, the availability of tens of thousands of clusters from X-ray and Sunyaev-Zeldovich surveys of the next generation, such as those promised by SPT and eROSITA, will bring the statistical uncertainties well below the percent level. In this case, the major limitation in constraining cosmological parameters is related to the systematic uncertainties in the relation between the observables, on which the cluster identification is based, and mass.

In order to control such systematics, two different approaches have been suggested. The first one, is based on the self-calibration method (Majumdar & Mohr (2003).) This assumes suitable parameterizations for the scaling relations between M and $L_X$. Taking advantage of the large number of clusters offered by future surveys, the parameters defining such relations are fitted along with the cosmological parameters. This approach provides much tighter constraints on cosmology if follow-up temperature measurements are available for a subsample of clusters. An alternative approach is based on finding a suitable observable, which is easy to measure and tightly correlated with mass. Kravtsov et al. (2006) suggested the quantity $Y_X=M_{gas}T_X$ to be tightly correlated with the total mass M (intrinsic scatter only 8%). Maughan (2007) found $Y_X$ to also be correlated with $L_X$ (intrinsic scatter 10%). Clearly, both approaches require in addition to the mere cluster detection a measurement of the temperature to keep the systematics that affect the determination of cosmological parameters under control. Thanks to its unique observational strategy, EDGE is capable of measuring temperatures for a significant fraction (about one quarter) of all clusters identified above the surveys' flux limits (Table 2.2.2). This allows to fully exploit the potential of the self-calibration method described above, and to compute the minimum-scatter mass proxy $Y_X$ for a large number of objects. We emphasize that no other planned or proposed X-ray mission has these characteristics. As such, EDGE will produce an ideal survey to exploit the potential of clusters of galaxies as tools for precision cosmology.

Relevance of EDGE results: In Fig. 2.2.12 we show the constraints on the $\Omega_m$-$\Omega_\Lambda$ plane which will be obtained by EDGE with both GRBs and cluster surveys, compared with present constraints from SNIa (Riess et al. 2004) and from the CMB (Spergel et al. 2006). Of course, future dedicated missions such as SNAP, if approved, will find and measure thousands of SNIa out to a limiting redshift of ~ 1.7. But while SNIa map out the low redshift range, corresponding to the Dark Energy dominance era, clusters and GRBs properly map the earlier matter-dominated era. In addition, as systematics dominate the uncertainties, only measurements with different probes will reduce the systematic errors. EDGE will contribute in two areas:

1. The present sample of GRBs with measured z and accurate estimate of $E_p$ will be extended from the present 20 events to ~80 events at the time of the launch of EDGE, mainly by Swift+GLAST and SVOM/ECLAIRS measurements. This extension will help to assess the relevance of the correlations used and to study the possible systematic effects, but it is still limited in terms of accuracy, homogeneity and number. With respect to these measurements, the sample of ~150 EDGE GRBs will provide a three times more accurate estimate of the cosmological parameters with a better understanding of the systematics (thanks to the homogeneity of the sample).

2. As for X-ray cluster surveys, eROSITA (launch in 2011) will detect ~$10^5$ clusters over a large portion of the sky (20,000 deg). However, the observational strategy and the characterization of the background prevents eROSITA from measuring temperatures for a useful subsample of clusters, with obvious implications on the possibility of controlling systematics in the estimate of cosmological parameters within the same survey.

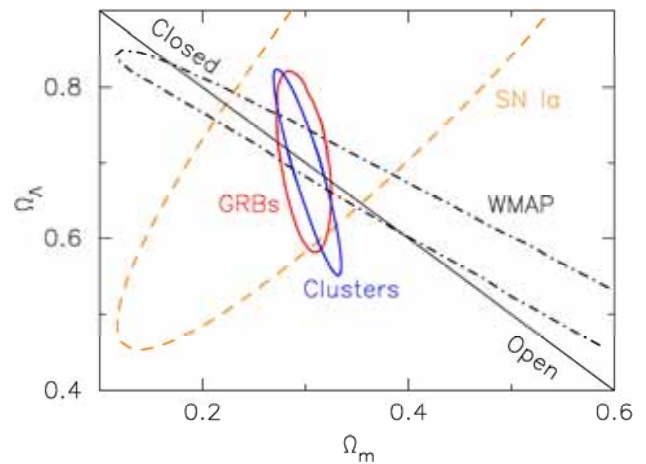

*Fig. 2.2.12 Confidence contours (99% confidence level) on the $\Omega_m$- $\Omega_\Lambda$ plane from different methods. GRB and Clusters are EDGE results, SNIa contours are from the sample by Riess et al. (2004) and CMB contours are from Spergel et al. (2007).*

### 2.3.2 Feedback processes

Feedback is a fundamental process in the evolution of the Universe. Supernovae heat and enrich their surroundings, the local ISM. The added winds or explosion shells of many stars or supernovae may produce galactic superwinds or superbubbles, powerful enough to enrich or heat the intergalactic space, and thereby affecting the evolution of the surrounding clusters and the WHIM. Outflows and inflows from the central black hole in galaxies and AGN, can provide the feedback mechanism necessary to link the growth of the Black Hole to the evolution of the host galaxy, as implied by the BH mass – bulge mass relation.

Supernova Remnants. The physical processes that are probably important for the properties of the WHIM, such as collisionless shocks, non-equilibrium ionization and heating, and possibly cosmic ray acceleration, have been studied extensively for Supernova Remnants (SNRs). However, even for SNRs we still lack a good physical model for collisionless shocks, a process taking place in very tenuous media where Coulomb collisions are inefficient. Nevertheless, the gas is hot, so some other heating mechanism, e.g. plasma waves, operates. A similar situation is likely to take place in the intergalactic shocks that form the WHIM. Another important aspect of SNR research is the study of fresh explosive nucleosynthesis products, especially in young SNRs. Also this has an important link with the cosmological research: the abundances of elements seen in clusters of galaxies, and to be expected in the WHIM, are caused by a mixture of Type Ia and core collapse supernova products.

EDGE has the necessary qualities to address the kinematics of different constituents of the remnant, as well as nucleosynthesis and collisionless shock physics. In spite of the modest spatial resolution of the WFS, EDGE is excellently suited for large SNRs. The kinematics of the SNR can be judged from Doppler shifts or broadening, whereas at the rims Doppler broadening will reveal the ion temperatures (see Fig. 2.2.13).

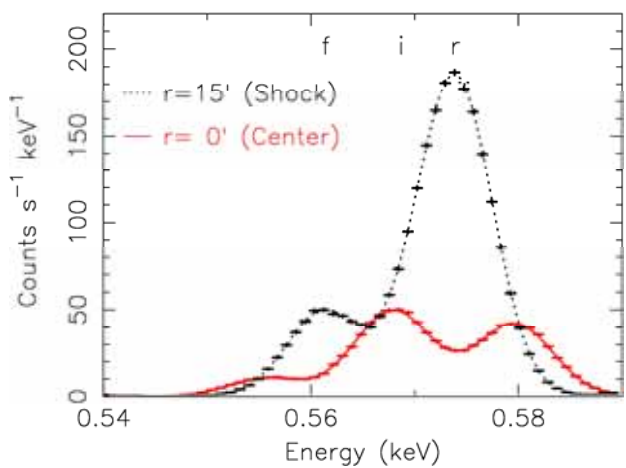

Fig. 2.2.13 Simulated spectrum of SN1006 with the WFS (100 ks) for a 3'x3'region at the center and at the shock front (15' from the center), The O VII triplet lines r, i and f are broadened to 9 eV due to the large ion temperature in both regions; at the center Doppler splitting of +6 eV and -6 eV between front and backside is seen.

Hot winds from our Galaxy and Starburst Galaxies. The local ISM of our galaxy is a complex and dynamical structure. It consists of dust, molecular clouds, cold and warm atomic gas, warm and hot ionized gas, all with different physical properties and spatial distributions. All these phases can be seen through X-ray absorption of background sources, and the hot ionized gas (~$10^6$ K) can be seen directly in continuum and line emission.

High spectral resolution observations at other wavelengths reveal the presence of high velocity clouds, and in several cases there is also hot gas associated with these structures (Kerp et al. 1999). Whether these high velocity clouds are the results of Galactic wind bubbles and fountains, tidal tails due to interactions with other galaxies (for example the Magellanic Clouds), or infalling pristine gas clouds associated with the WHIM is unclear. X-ray absorption studies of the ISM have become possible recently using the grating spectrometers of *Chandra* and *XMM-Newton*. In X-rays, all atoms contribute to the continuum opacity, regardless in which phase of the ISM they reside. With high spectral resolution it is possible to distinguish ions of different ionization stage as well as whether atoms are free or bound in, for example, dust grains. Answering questions such as whether the strong O VII absorption lines at z=0 are due to the WHIM in the Local Group of galaxies or the ISM of our Galaxy depends critically upon the availability of high quality spectra, and preferentially both emission and absorption spectra of the same region.

Given the present observational plan, we expect hundreds of pointings (those associated with GRB follow up) in random directions of the sky, which result in high quality spectra and eventually allow to map chemical abundances and velocity fields of different phases of the ISM. Absorption studies of bright Galactic point sources will reveal the total column density of the line of sight while the chemical shifts due to binding in molecules, well separated in energy by EDGE from the atomic phase, allow us to determine the condensate state of the medium. The combination of absorption and emission measurements both yields density and absolute size. EDGE is the first instrument that can do this.

The dynamics of the galactic hot ISM can be observed more dramatically in nearby starburst galaxies, where metal-rich gas is flowing out due to an extreme supernova activity in the nucleus.

AGN outflows. AGN X-ray spectra are rich in discrete spectral features, both in emission and absorption. Spectral diagnostics have revealed that a large amount of matter is actually outflowing with velocities ranging from hundreds to several thousands of km/s, with the higher velocity clouds having usually higher ionization states. The low velocity outflows (also known as "warm absorbers") are very common, being present in about half the sources in the local Universe, while the high velocity clouds seem to be a characteristic of sources accreting at high rates. The warm absorber outflows are likely to originate from the dusty torus expected to surround the nuclear black hole. The energy they carry is <1% of the AGN bolometric luminosity but their mass outflow rate is often greater than the mass accretion rate powering the AGN. Over an AGN lifetime, the outflow rate is sufficient to supply, for example, all the hot ISM in a Seyfert galaxy bulge (Blustin et al. 2005). AGN outflows will then affect the relation between the Super Massive Black Holes (SMBHs) and their host galactic bulges, a link that will be further investigated in the next Section.

### 2.3.3 Active Galactic Nuclei

There is a close link between the growth of SMBHs and the formation and evolution of the host galaxy (Ferrarese & Merritt 2000). In order to understand this co-evolution we need to study the processes which make SMBH grow and which determine their accretion rates and accretion efficiency that are likely to be affected by AGN outflows. Numerical simulations show that SMBHs grow quite rapidly, but remain buried in gas and dust for long periods. The turning-on and evolution of the accreting SMBHs and their importance in the context of galaxy formation and evolution are best studied using an appropriate census of the AGN population.

The EDGE Deep Surveys (1-2 Ms exposure) will provide a catalogue of more than 60,000 AGN, down to a limiting flux of $1.5 \ 10^{-16}$ erg s$^{-1}$ (three times the Chandra Deep Field South limiting flux). The surface density is more than three times larger than obtained with the XMM-COSMOS survey (Hasinger et al. 2006). The EDGE Wide Survey (50 ks exposure) covers nearly 1% of the sky; it contains 200,000 detectable AGN. The huge EDGE catalogue will enable a very detailed study of the evolution of the luminosity function in different luminosity bins at redshifts up to z=6. Also, the faint end slope of the luminosity function at high redshift and the space density of very high redshift AGN will be directly measured. Recent models for the X-ray background (Gilli, Comastri & Hasinger 2007) predict ~3000 AGN at z>4 and ~1000 at z>6 detectable in the Deep Survey with the WFI. X-ray spectroscopy with the WFS will provide a powerful tool to uncover high z quasars and directly measure redshifts from the 6.4 keV iron line. The WFS allows the detection of narrow iron lines with rest frame EW down to 150 eV at redshifts up to 10. At a 0.5-2 keV flux level of $10^{-14}$ erg cm$^{-2}$ s$^{-1}$, above our confusion limit, we expect to measure spectra of 30,000 galaxies in the Wide Survey, 200 of them at z>4 and 30 at z>6.

The EDGE survey probes the AGN spatial distribution up to scales of ~50 Mpc (approximately 3° of the sky at redshifts 0.5-5) over a large range of luminosities. The precise measurement of such evolution and its dependence upon luminosity will have important consequences on our understanding of the formation process of supermassive black holes, their co-evolution with galactic bulges and dark matter haloes, the turning-on of the AGN phase and of their life times.

### 2.3.4 Physics and progenitors of GRBs

In spite of our understanding of long GRBs as the explosions of massive stars, our knowledge about the various subclasses of GRBs, their progenitors, and actual physical processes of GRB formation is still quite limited. For the origin of short GRBs, merging of binary neutron stars or a neutron star and a black hole has been proposed. There are X-ray flashes, variants of GRBs with peak spectral energies in the X-ray band. There could be also a numerous class of sub-luminous explosions, with less relativistic ejecta, that would outnumber GRBs by orders of magnitude (Soderberg et al. 2006). EDGE will allow us to observe many more XRFs and short GRBs. Detailed afterglow studies help to reveal their origins because the afterglow retains a memory of the environment in which the explosion takes place. The association of long GRBs with peculiar type Ibc SNe implies a metal rich environment and a wind density profile. The presence of substantial absorption in the X-ray spectra has already been noted. If this absorbing material is located close to the GRB site, the low energy cut-off is expected to fade as a result of the progressive ionization by GRB high energy photons. Tracking the evolution of the absorber via the enhanced statistics and spectral resolution of EDGE allows an unprecedented characterization of the circumburst medium (density, metallicity and distance of the material from the GRB).

GRBs also provide a test of physics under the most extreme conditions. GRBs consist of a newly formed black hole, or possibly a short-lived magnetar. This central engine converts a substantial fraction of its energy into a jet traveling at the highest Lorentz factors observed in the Universe. Small variations within the jet itself and its interaction with the external medium convert bulk kinetic energy into high energy radiation through relativistic shocks and particle acceleration (the fireball). We require the broad bandpass of EDGE to clarify the origin of the correlations linking the total electromagnetic energy with its peak frequency, to constrain spectral evolution as the peak moves with time, and hence to test prompt emission mechanisms. Previous observations also show late X-ray flares and a plateau phase which strongly imply a central engine that is still alive and pumping substantial power many hours after the initial outburst. The large area of the EDGE telescope strongly constrains, through monitoring short timescale variability, the origin of these puzzling phenomena.

### 2.3.5 The densest matter in the Universe

The densest matter in the observable Universe is within neutron stars (NSs), with densities exceeding that of nucleonic matter ($\rho > 3 \times 10^{14}$ g cm$^{-3}$). They are unique laboratories for the study of fundamental particles and of the strong interaction under high $\rho$ and low $T$ conditions.

They may contain quarks unconfined to hadrons. The equation of state (EOS) determines the relationship between mass $M$ and radius $R$ and depends on the composition of the NS. EDGE, through its monitoring, quick slewing and high spectral resolution capabilities, can make a unique contribution to constraining the EOS by following up rare but energetic thermonuclear X-ray bursts on NSs in our galaxy.

NSs in low-mass X-ray binaries accrete matter from the atmosphere of a neighboring non-compact star, and burn that matter through unstable thermonuclear shell flashes on the surface. These flashes yield the most luminous thermal radiation from NS surfaces, and provide very good opportunities for constraining the $M/R$ ratio through the measurement of gravitationally redshifted narrow spectral lines and edges. Initial tentative evidence for lines with redshift 0.35 was obtained from X-ray burst spectra of EXO 0748-676 by Cottam et al. (2002). The derived value for $M/R = 0.153$ $M_\odot$/km is consistent with most modern NS EOSs composed of normal matter. Much larger EWs are expected in X-ray bursts characterized by long-duration photospheric radius expansion (PRE, Weinberg et al. 2006), and possibly from so-called superbursts (Cornelisse et al. 2000; Cumming & Bildsten 2001) that liberate $10^3$ times more photons from the NS surface than ordinary bursts. Their durations, typically tens of minutes for the former and hours for the latter, are well within the reach of the EDGE capabilities. We need 5-10 bursts with gravitational redshift to constrain the EOS (see Fig. 2.2.14). The frequency of suitable bursts is once every few to hundreds of days for each of about 20 sources. The expected number of bursts per year detected by the WFM is 20.

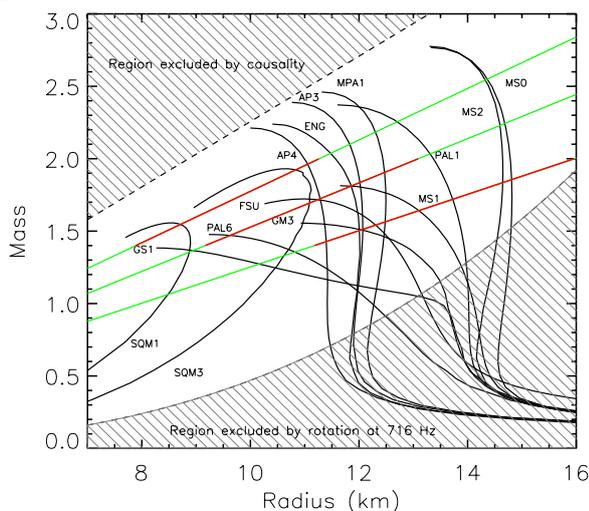

Fig. 2.2.14: *mass-radius relations for various equations of state of dense matter including standard nucleonic matter and strange quark matter (see Lattimer and Prakash 2001). To demonstrate the potential of EDGE, we show the constraints obtained from the measurement of redshifts of three bursters (green lines); the red parts highlight a likely NS mass between 1.4 and 2.0 $M_\odot$).*

### 2.3.6 Violent accretion on compact objects

Flaring behaviour of compact sources, black holes and neutron stars, on time scales of hours to minutes is today a rather established, but poorly studied phenomenon. Strong outbursts have been found in BH systems, which may indicate violent activity of the inner accretion disk (e.g. feeding of material into jets), fluctuations of direct wind accretion, or the presence of thick, possibly clumpy absorbers or outflows. For classical nova outbursts as well as for persistent sources much is left to understand regarding the jet formation mechanism. The jet can possibly be formed in the soft to hard state transition typically observed in bright BH transients towards the end of the outburst. The presence of ionized outflows, likely to be related to the jets, has been demonstrated by *Chandra* as blueshifted absorption lines for a few objects. It is not yet clear whether the optically thick, ionized flow is related to a jet, or a disk wind resembling the warm absorbers which are present in Seyfert galaxies.

The X-ray properties of different kinds of sources depend also on the type of the donor star (low mass or high mass star, supergiant or Be star), and reveal the evolutionary history of the binary system. Supergiant Fast X-ray Transients (SFXT) are a new class of sources, found using the INTEGRAL monitoring of the Galactic plane These are high-mass X-ray binaries with blue supergiant companions (O-type or early-type supergiants). They display extremely short hard X-ray outbursts, with a typical duration of few hours. The rapidly growing number of SFXTs demonstrates that they may be a dominant population of X-ray binaries, born with two very massive stars. They could be the progenitors of double neutron star binaries or of the neutron star-black hole binaries, and thus SFXTs could be the origin of possible NS/NS or NS/BH mergers. The study of SFXTs can play a crucial role in the search for sources of gravitational waves and for understanding the origin of short gamma-ray bursts.

EDGE provides a unique opportunity to study fast transient episodes from SFXTs. Rapid response to X-ray transient events and the capability of high resolution spectroscopy are the keys to probe the extreme relativistic regime of the inner regions in BH accretion disks and the physical conditions at the neutron star surfaces.

### 2.3.7 Stars

EDGE, with its deep observations and relatively large field-of-view will contribute significantly to the study of star forming regions, stellar populations and the physics of stellar flares.

Complementary to other instruments, such as Spitzer, ALMA and 2nd generation VLTI instruments, the X-ray window is important for understanding the accretion and outflow processes including the role of magnetic fields. These are key processes in the formation of stars and of planetary disks (e.g. Feigelson et al., 2007). A large survey of stars gives us a unique opportunity to understand the Class III population and to compare this with the class 0, I and II populations as determined by Spitzer.

Coronal sources are known to produce flaring X-ray emission with released soft X-ray fluence >10[37] erg and on time scales up to a week. However the understanding of the evolution of the hot plasma and the role of impulsive or continuous heating has a limited observational base. Using known flaring objects (e.g. Algol, AR Lac) and the fast trigger capability of EDGE we can study these processes in great detail.

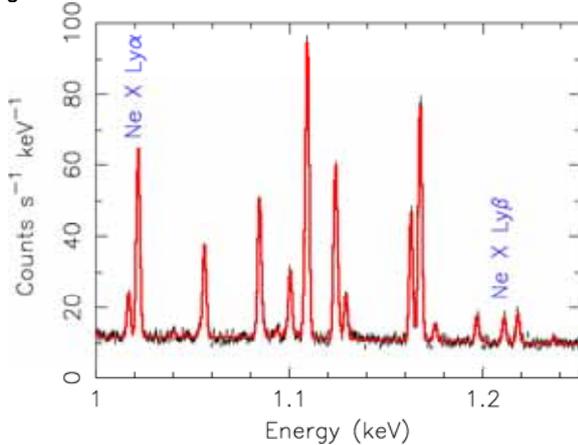

*Fig. 2.2.15 WFS spectrum of a flare of Algol (exposure time 30 ks, typical 0.2-10 keV flux (1% of giant flare) of 10[-10] erg s[-1]). Most lines are due to Fe XX- XXIV except the two labeled Ne lines.*

### 2.3.8   The Solar System

Some of the extreme processes that take place in planetary magnetospheres, in particular those leading to aurorae production, involve highly energetic plasmas and interaction/emission mechanisms that are widespread in astrophysics, and of which planets are our 'next door' examples.

Different processes are known to produce X-rays in solar system bodies. Most of the soft X-rays lines (predominantly O VII and O VIII) observed in objects like Jupiter, Mars, comets, and the Earth, are produced by interaction of highly ionized ions of the solar wind, such as C, N, O with atmospheric gas through charge exchange (SWCX). Particularly intriguing is the origin of the soft X-ray lines observed in the aurorae of Jupiter (Branduardi-Raymont et al. 2007). Determining the ion species via high resolution spectroscopy with EDGE will tell us whether the ions originate from the solar wind penetrating the magnetosphere, or from the volcanoes of Jupiter's satellite Io. Spectroscopic studies with EDGE of SWCX, (the emitted power, spatial distribution and ionic species involved), will be important in furthering our understanding of the solar wind interactions throughout the heliosphere and in probing the atmospheres of solar system objects, including the Earth. Since this component is bright and likely to be very extended (Fujimoto et al. 2007), it has a significant impact on the derivation of precise line intensities for the ISM and the WHIM. EDGE observations provide essential data in a serendipitous manner, resolving lines of different origins, also using that charge exchange leads to different line ratios than found in thermal plasma such as the ISM.

### 2.3.9   Search for light dark matter

The quest for Dark Matter (DM) particles is one of the major challenges in Particle Physics and Cosmology. Its nature greatly affects the structure formation. Recently, a simple and natural extension of the MSM (Minimum Standard Model) was suggested. This model explains all observed data for neutrino oscillations and baryogenesis in the Universe. At the same time it provides a warm DM particle (Asaka et al. 2005) consistent with present evidence of bottom-up structure formation. However, on galactic and extragalactic scales cold and warm DM models can be distinguished experimentally, This warm DM candidate is often called *sterile neutrino* to distinguish it from the usual *active (left-handed)* neutrinos. Sterile neutrinos interact with other matter only via mixing with active neutrinos. Sterile neutrinos with a mass in the keV range can be produced in the early Universe in several different ways (Shaposhnikov & Tkachev 2006). These particles can be searched for in X-rays, because they decay into an active neutrino and a photon with energy $E_\gamma = M_s/2$. From the Ly-$\alpha$ forest a lower bound to the mass of 1 keV can be derived, thus leaving a window of parameters that can be probed by the EDGE Wide-Field Spectrometer. The expected decay flux is:

$$F_{DM} = 6.38 \ (M_{DM}^{fov}/10^{10}M_\odot)(D_L/Mpc)^{-2}$$
$$(M_s/keV)^5 \sin^2(2\theta) \ keV \ cm^{-2} \ s^{-1}$$

where $M_{DM}^{fov}$ is the mass of DM within the telescope's FoV and $D_L$ is the luminosity distance to the object. Sterile neutrino interaction with the active counterparts is parametrized by the *mixing angle $\theta$*. The width of the decay line is determined by Doppler broadening: $\Delta E/E \sim 10^{-4}$ for a typical dwarf spheroidal to $10^{-2}$ for a cluster of galaxies. A number of searches for this line have been conducted (e.g. Boyarsky et al., 2006), setting an upper limit on the mixing angle as function of energy. At 1 keV with typical parameter values for the DM particle, the expected line flux is <0.4 photons s[-1] for the EDGE grasp. The sensitivity to line detection against a known continuum spectrum increases with the square root of the GRASP and with the spectral resolution as $\Delta E^{-1/2}$, thus making EDGE the perfect mission to conduct this search. As shown by Boyarsky et al. (2006) the signal from al-most all nearby objects (dwarf galaxies, Milky Way, large elliptical galaxies, clusters of galaxies) provide comparable DM decay signals. In this sense, the search can be carried out serendipitously in most of the observations from EDGE.

## 2.3.10  Gravitational waves

Coalescing supermassive binary black holes are primary sources of gravitational waves (GWs) detectable throughout the entire Universe by *LISA* (Bender et al. 1994). These events occur during the hierarchical assembly of galaxies when (pre-)galactic substructures undergo collisions and mergers. *LISA* is expected to detect ~35 coalescences in three years of operation out to z<10 with S/N>100 (Sesana et al. 2004). GWs alone will provide an accurate measure of the source luminosity distance $D_L$ (Vecchio 2004), with a relative accuracy of ~1%. A GW source coupled with an electro magnetic-counterpart, providing the redshift, could constitute a "standard siren". Identification of a few standard sirens could become our best observational probe of the geometry of the space-time and of the expansion history of the Universe, provided that a good modeling of weak lensing is performed (Holz & Hughes 2005). X-ray "afterglows" could be produced by the resumption of accretion on the merged massive BH, which is expected to occur within a few years after the moment of coalescence and to activate X-ray emission (Milosavljevic & Phinney 2005). These afterglows should be searched for in a FoV as wide as the *LISA* error box, which can be 1 deg². Observational requirements are high sensitivity and large FoV. EDGE is unique in achieving the required performances.

The procedure to identify the candidate will not be trivial. GW detections from *LISA* provide the location and tell in advance the likely X-ray flux of the source. Above S>5 10⁻¹⁶ erg s⁻¹ cm⁻² Dotti et al. (2006) predict at least 3 sources, 1 with S>10⁻¹⁵ erg s⁻¹ cm⁻² (in ~3 yr from coalescence). At these flux levels we expect hundreds of AGNs, which are known to be notoriously variable, in the field of view. Using the delayed (3 years) turn on expected for the SBBH afterglow, monitoring the area on the sky with the GW source (100 ks observations, commencing right after the GW trigger) we will separate the variable AGN sources from the GW source. Clearly the extended mission duration would enhance the capability.

Since GRBs are collimated, there can be 10 times more events escaping detection with EM waves. Advanced LIGO and Virgo observations will detect Gravitational Waves from GRBs, which is more isotropic emission, and give trigger signals for EDGE to monitor X-ray emission. This enables us to detect all of the GRB events in the Universe and to study X-ray properties of misdirected GRBs.

# 3  Mission profile

The measurement of low surface brightness objects and weak emission lines or absorption features is extremely challenging and requires a well balanced combination of instrumental capabilities including:

- *High resolution X-ray imaging spectroscopy*
- *High angular resolution with high contrast and low background*
- *Wide sky monitoring to find GRBs*
- *Fast autonomous pointing towards transient sources*
- *Wide field coverage (~1 deg²) with long observations (~Ms)*

Employing recent developments in detector and mirror technology it is now possible to perform these measurements. The requirements for the various key science topics are listed in Table 3.1. We have also specified realistic goals (see Table 4.1) for which the main effort will be to improve the energy resolution and extend the energy band of the spectrometer. The requirements are achieved by a suite of 4 instruments:

- a Wide-Field Spectrometer (WFS) with high spectral resolution
- a Wide-Field Imager (WFI) with high angular resolution
- a Wide-Field Monitor (WFM) which monitors a significant part of the sky (2.5 sr) and triggers the fast re-pointing towards GRBs
- a GRB detector (GRBD) which extends the energy range of the WFM. This extension allows for improved triggering of the WFM and some very exciting additional results (see Section 2.3.1)

In order to verify that the instrument performance can meet the demands of the science case, we have used fully representative instrument responses for all the instruments.

## 3.1  Launch and orbit

The required low background dictates a low equatorial earth orbit (LEO). While this has a number of additional advantages (VEGA launcher, S-band), it results in additional constraints on the thermal design due to the vicinity of the earth. Initial analysis has demonstrated that this is feasible, and a low earth orbit is currently also planned for the Japanese NeXT mission.

*Table 3.1:  overview of the science requirements (for goals see Table 4.1, 4.2, 4.3, 4.4)*

| Science | Effective area [cm²] | Energy range [keV] | Angular resolution | Field of view [deg²] | Spectral resolution | Instrument |
|---|---|---|---|---|---|---|
| Missing baryons in absorption | 1000 @ 0.5 keV | 0.2 – 1 | n/a | n/a | 3 eV @ 0.5 keV | WFS+WFM |
| Missing baryons in emission | 1000 @ 0.5 keV | 0.2 – 1 | 4' | 0.7 x 0.7 | 3 eV @ 0.5 keV | WFS+WFI |
| Cluster physics | 500 @ 1 keV | 0.3 – 4 | 15" | 1.4 ⌀ | 80 eV @ 1 keV | WFI+WFS |
| Cluster formation and evolution | 500 @ 1 keV | 0.3 – 4 | 15" | 1.4 ⌀ | 80 eV @ 1 keV | WFI+WFS |
| Metal enrichment | 1000 @ 0.5 keV | 0.2 – 2 | n/a | n/a | 3 eV @ 0.5 keV | WFS+WFM |
| Dark ages | 1000 @ 0.5 keV | 0.2 – 2 | n/a | n/a | 3 eV @ 0.5 keV | WFS+WFM |

80 bursts per year with fluence > $10^{-6}$ erg cm$^{-2}$ in the 15 – 150 keV energy band

*Table 3.2: overview of instrument requirements*

| Instrument | Aeff [cm²] | E-range [keV] | Angular resolution | Field of View [deg²] | spectral resolution |
|---|---|---|---|---|---|
| Wide Field Spectrometer (WFS) | 1000 @ 0.6 keV | 0.2 – 2.2 | 3' | 0.7 x 0.7 | 3 eV @ 0.5 keV |
| Wide Field Imager (WFI) | 500 @ 1 keV | 0.2 – 5 | 15" | 1.4 ⌀ | 80 eV @1 keV |
| Wide Field Monitor (WFM) | 500 @ 50 keV[1] | 8 – 200 | 35'[2] | 2.5 sr | 3% @ 100 keV |
| Gamma-Ray Burst Detector (GRBD) | 800 @ 600 keV | 25-2500 | n/a | 3 sr | 20% @ 100 keV |

[1] Averaged over 2 sr.
[2] More critical is the location accuracy of 4'

The instruments will be accommodated on a single three-axis stabilized spacecraft which provides all necessary services. A noteworthy property of EDGE is the fast and autonomous repointing capability. Following the onboard detection of a GRB by the WFMonitor, the satellite will be able to slew quickly and autonomously to its position. The accuracy of the new pointing will be 4' which is compatible with the sector of the WFSpectrometer which can handle high count rates. In addition to the usual ground contact during the passage of the satellite over the ground station, a continuous capability to transfer GRB related information (positions, spectra, etc.) to the ground will be implemented. This system can also be used to upload TOOs to the satellite with a reaction time on the order of minutes (compared to the usual reaction time of hours).

The satellite is compatible with the VEGA launcher which has the capacity to bring a payload of 2300 kg into an equatorial orbit of 500 - 600 km. Significant effort has been applied to optimizing the scientific payload within the available volume since there is a need for 11 m² of solar panels, in addition to the instruments themselves. The result is illustrated in Fig. 3.1 where we display the dynamic envelope of the VEGA launcher and the satellite in its launch configuration. The main satellite requirements are given in Table 3.3. As indicated, we have selected a nominal duration of 3 years, sufficient to achieve the main science goals of EDGE. The extended lifetime of the mission (5 years) can be achieved with a reasonable level of redundancy and reliability, consistent with the envelope of an M-class mission.

The proposed satellite has a remarkable heritage in the NASA SWIFT mission and ESA GOCE mission. In addition to these missions, the satellite has heritage in other well known high energy astrophysics missions in Low Earth Orbit such as BeppoSAX, Suzaku and the recently launched AGILE.

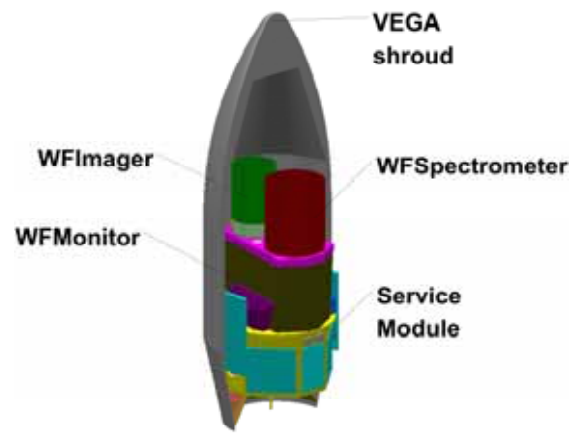

*Fig 3.1    Satellite in its launch configuration*

Table 3.3  Main satellite characteristics

| Parameter | Requirement | Conformance |
|---|---|---|
| Orbit | LEO, ≤ 5 ° inclination | Compliant |
| Pointing | 3 axis stabilized | better than 2', knowledge < 1" |
| Re-pointing | 60° in 60 s | In 60 s for 80% of the slews, in 45 s for 50% of slews feasible |
| Lifetime | 3 years | 5 years, longer lifetime feasible if funding permits |
| Mass | 1932 Kg | Launch capability: 2300 kg |
| power generation | 3800 W | Includes battery charge and peak power during repointing |
| Telemetry | 3.8 Mbps continuous coverage | compliant ORBCOMM system for small volumes of data |

## 3.2   Operations

*Ground segment*

The selected orbit allows for ground contact every orbit. With appropriate data selection and compression, already achieved on current satellites, a TM rate of 3.8 Mbps is sufficient. This can be provided by the enhanced S-band capability, which will be available in the 2017 time-frame. We have selected the Korou ground station (15 m dish), but other ground stations (e.g. Malindi) can be considered as well. For continuous communication (GRB position downlink, TOO uplink) the commercial available ORBCOMM system will be the baseline. This system is being used for the Italian AGILE mission. The down link time for small data volumes (<250 bytes) varies between <5 min (30% of the messages) to <20 min (70% of the messages). Clearly a shorter response time, either available through further improvements of this system, or through the US TDRS system, or through a network of small antennas along the equator, is of great importance for optical and IR follow up measurements of GRBs.

The required onboard capability to point autonomously at a GRB will include implementation of viewing constraints during slews and thus reduces the load on the ground system considerably. This system will also be used for TOOs and normal slews, avoiding a significant load on the mission planning group.

*Observing efficiency and sky visibility*

Using the repointing capability, it is possible to optimize the observing efficiency considerably. If the viewing of the science target is blocked (by the Earth, Moon,…), another target will be selected. It has been demonstrated by the SWIFT satellite that the observing efficiency can be 70% using this capability and we assume a similar efficiency.

The Attitude and Orbit Control System allows for:

- *fast autonomous follow-up observations towards transient and bursting sources triggered onboard.* Efficient trigger criteria will be set taking into account the characteristics of different events (correlations between energy, duration, and position). This improves, among other things, the capability to find high redshift GRBs
- *smart pointing to maximize the observing efficiency.* Once a pre-planned target is being occulted by Earth, the spacecraft will slew, at a lower rate, to another pre-planned target
- *pointing to a TOO target,* uploaded from the ground through the ORBCOMM system

For optical follow up of GRBs the preferred pointing direction of the WFM is in the anti-sun direction. In the current design this is partially achieved due to the large field-of-view of the WFMonitors.

All slews are checked onboard for observing constraints along the entire slew path, reducing the need for operator interventions. Each target is assigned a priority (by default GRBs and TOO are executed immediately). Predefined locations near the poles will be used as safe mode pointing positions. During one orbit, 70% of the sky will be accessible, with the largest constraints imposed by the Sun and Earth limb avoidance angles of 45° and 10°, respectively. Direct sunlight on the radiators has to be avoided but these constraints can potentially be relaxed using heat diodes.

Based on existing heritage, no critical issues have been identified for the mission profile. Optimization of the satellite design (solar panels, radiators), with respect to the Sun aspect angle, will be important during the next phase since this could enlarge the possible viewing directions.

## 3.3  Observation program

The observation program includes two parts:

*a)    the core program.* This is ~80% of the observing time during the first three years. The driving science topics will be addressed in this time. It includes deep observations of the WHIM and clusters (typically 1 Ms per observation), and GRB follow up measurements. These data will be made public periodically (once per month). Note that the long observations for the core program will be used for the auxiliary science as well.

b)   *a guest observer (GO) program.* ~20% of the observing time which will be assigned following normal ESA guest observer practice.

In Table 3.4.a we show the planned observing times for the core program. A significant part of the auxiliary science will be obtained as serendipitous data from the core program. This is indicated in table 3.4.b, and some tentative observing times are assigned to the other topics. Clearly, there is room for a competitive GO program.

*Table 3.4.a: Core observing program in 3 years*

| Science driver | Observing time # [Ms] | id[1] |
|---|---|---|
| Missing baryons in absorption (240 bursts) | 240 x 0.05 | A |
| Missing baryons in emission (3.5 x 2.1 deg$^2$ + 3 follow up) | (15+3) x1 | B |
| Cluster physics: 9 clusters + | 8.5 | C |
| 9 background observations | 8.5 | D |
| Cluster formation (100 deg$^2$) and evolution | 100 x 0.05 and A + B + D | E |
| From local to the Dark Universe | | A |
| Total | 52 | |

1) identifiers used to indicate the parts of the core program which will be used for auxiliary science as well

*Table 3.4.b: Auxiliary science program (total 13 Ms available)*

| Science driver | Observing time # [Ms] |
|---|---|
| Cosmological parameters (GRBs + clusters) | A + (D+E) |
| Feed back in action | 4 |
| GRB physics and progenitors | A |
| EOS of densest matter | 1 |
| compact objects | 2 |
| AGN survey | B + D + E |
| Star survey and variability | B + D + E + 1 |
| Search for light dark matter | D + 1 |
| gravitational waves | 3 |
| Solar system | 1 |
| Total auxiliary science | 13 |

Target selection for the core program will be done about one year before launch in consultation with the community (see section 6.1). Especially for the large (contiguous) fields, it is important to be sufficiently close to the polar regions (for year round visibility), to have a relatively low $N_H$ (< 2 $10^{20}$ cm$^{-2}$) to reduce the loss due to absorption, to avoid large scale Galactic structures (in the ROSAT ¼ and ¾ bands), and to have a good overlap with available surveys.

### 3.4   EDGE and other facilities

EDGE is highly complementary to existing and planned X-ray missions. Its high spectral resolution and wide field-of-view, supplemented by the high angular resolution of a more classical X-ray telescope, makes it unique for the detection of weak emission and absorption features in extended sources. This is illustrated in Fig. 3.2 where the GRASP (effective area x solid angle) and energy resolution are plotted for current and planned missions. By a GRASP which is about two orders of magnitude better than future observatories, or by an energy resolution which is also about two orders better than current and planned missions, EDGE will be significantly more sensitive for weak line detection of extended sources than any of these missions.

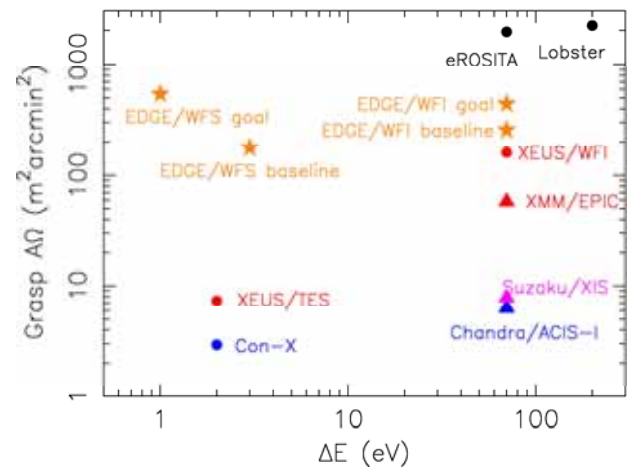

*Fig. 3.2 GRASP and energy resolution @ 0.5 keV for current and planned missions. The GRASP of NeXT is well below Con-X.*

A second unique feature of EDGE is its ability to acquire deep exposures of a large field-of-view with good and constant spatial resolution, by placing this suite of instruments in a low background orbit. This results in flux limits comparable to, or better than, many performed and planned surveys. This is shown in Fig. 3.3, where flux limits and sky area of performed surveys and the planned eRosita survey are given. The good PSF over the full WFI FoV, combined with its low instrumental background (due to the LEO) and small f number, will result in sensitive surveys over a large solid angle. At this sensitivity the 15'' instrument PSF is just consistent with the confusion limit.

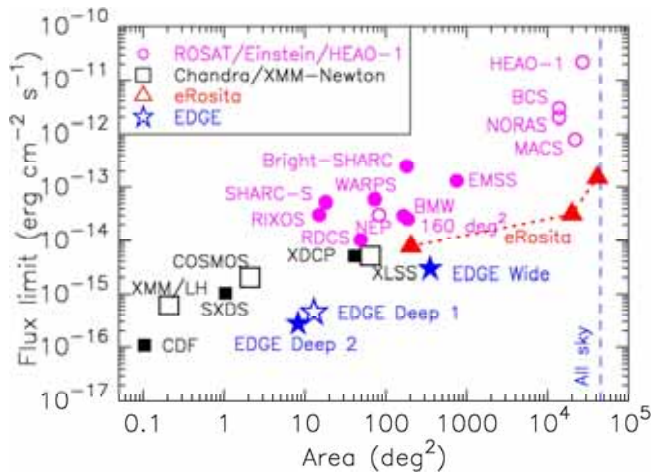

Fig. 3.3 Solid angles and flux limits of X-ray cluster surveys carried out over the last 2 decades or planned in the future (eROSITA).

A third feature of EDGE is the fast repointing of the satellite. This enables for the first time high spectral resolution observations of bursting sources. It allows us to study the properties of the GRBs and their local environment.

# 4   Payload

A single instrument is not able to provide simultaneously good spectral resolution, good angular resolution, low background, and triggers for fast repointing. Hence, we have designed a set of four complementary instruments which, together, meet the science goals of the mission:

- The *Wide-Field Spectrometer* (WFS) which has a cryogenic detector (100 mK) with high spectral resolution and a modest PSF. The WFS focal length is very short to obtain a large field-of-view for a very small detector. The required modest angular resolution (4') can be provided using light-weight foil optics.

- The *Wide-Field Imager* (WFI) which has a mirror with a good HPD (<15") over a relatively large field-of-view, a moderate focal length (2.75 m), and a large cooled CCD detector.

- The *Wide-Field Monitor* (WFM, 2 units) which is a coded mask instrument with a hard X-ray CdZnTe detector (8–200 keV) and a very large field-of-view (2.5 sr). It provides the capability to detect and localize GRBs and other transient events.

- The *Gamma-Ray Burst Detector* (GRBD 2 units) which consists of crystal scintillators and extends the energy range beyond 1 MeV. This provides extra triggering for GRB detection and is needed to determine the peak energy of GRBs with an accuracy of 10% or better. These detectors have no imaging capability.

For each of the four instruments we will discuss the key requirements, describe the current status in some detail (baseline), and indicate realistic goals which can be expected based on ongoing development programs with our partners. Some generic aspects, including required budgets and calibrations, are presented at the end of this section.

## 4.1   Wide-Field Spectrometer (WFS)

High resolution spectrometry is a major part of EDGE, enabling key measurements related to GRBs, the WHIM, and clusters. In order to meet the scientific requirements one needs the capability to disentangle weak emission and absorption features from prominent foreground emission and the instrumental background. Current advances in cryogenic detectors with imaging capability and a spectral resolution of a few eV at 500 eV make this feasible for the first time.

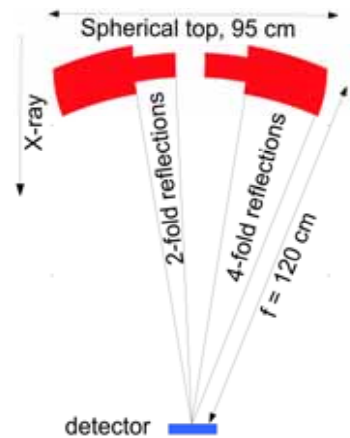

Fig 4.1   Schematic Instrument design of the WFS.

The instrument design is shown in Fig. 4.1 and includes a telescope with a conical shape with 2 and 4 fold reflections for the inner and outer shells respectively and a TES array in a cryogen-free cooler, which will be cooled to 50 mK. After selecting a very short focal length (1.2 meter), the baseline FoV is 0.7 x 0.7 deg$^2$ for a 32 x 30 array of 0.5 x 0.5 mm$^2$ pixels. The standard pixel has energy resolution <3 eV, but smaller pixels with lower heat capacitance can achieve even better performance (a factor of 9 in heat capacitance translates into a factor of 3 in the energy resolution). Lithographic technologies allow for a hybrid device, of which a subset of the pixels has significantly better energy resolution. In addition, a small out-of-focus section (64 pixels) can be included to avoid pile-up for observations of strong sources, like gamma-ray bursts in cases for which the count rate exceeds 200 counts/pixel (if the angular resolution is better than 4'). Operation at 50 mK with 0.4 µW cooling power can be obtained by a cryogenic-free cooler that requires 203 kg and 424 W.

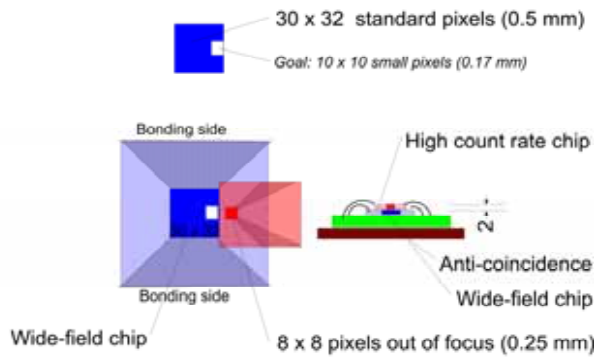

Fig 4.2 the detector design including the out-of-focus sector (red) to handle bright GRBs.

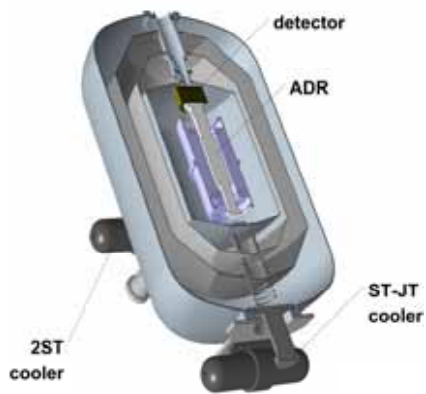

Fig. 4.3. The design of the cooler is shown to illustrate the level of detail used to estimate the instrument performance and resources. Alternative designs are also under study.

The instrument clearly meets the required performance, as shown in Fig. 4.4 and 4.5. An energy resolution of <2 eV has been demonstrated by the EURECA team on single pixels at the Bessy synchrotron over an energy range between 0.15 and 2 keV (the Cu absorber in these detectors needs to be replaced by a Cu/Bi absorber to enlarge the pixels). A flight qualified cooler is being developed for the Japanese NeXT mission, due for launch around 2013. The principle of four-fold reflections has been verified by one of our partners (Nagoya University) and is an extension of available mirror technology. With a positioning error of the shells of ±8 µm, already achieved on the Japanese Suzaku mission, the half power diameter is already 3.7', which is slightly better than required. The detection efficiency at low energy is limited by four entrance filters (50 nm parylene + 20 nm Al on a Si support grid) at the different temperature stages of the cryostat. The mirror temperature is controlled using a long thermal baffle and 125 W heating power. No thermal blanket is required on the mirror to keep the mirror at its operational temperature.

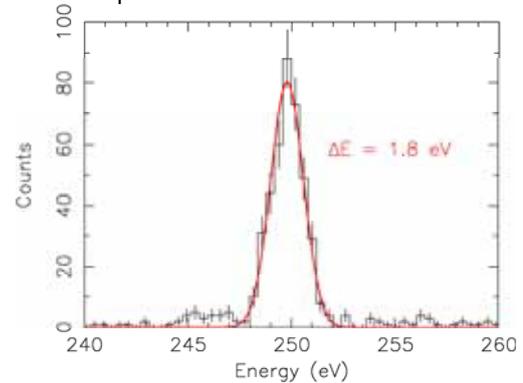

Fig. 4.4 Measured energy resolution of the WFS detector at Bessy (tail to low energies is due to test setup aspects).

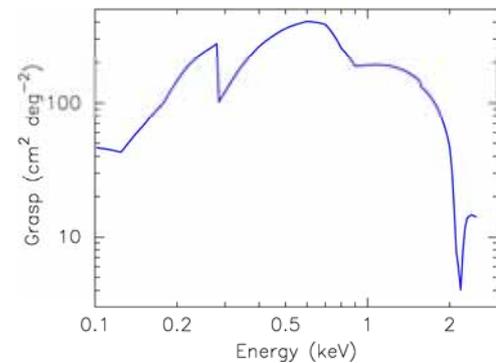

Fig. 4.5. WFS GRASP including mirror vignetting, detector filters and detector efficiency.

Table 4.1 WFS requirements, baseline and goals

| Parameter | Requirement[1] | Baseline[1] | goal[1] |
|---|---|---|---|
| Resolution at 0.5 keV [eV] | 3 | 3 | 1 |
| Field of View [deg²] | 0.7 x 0.7 | 0.7 x 0.7 | 0.8 x 0.8 |
| Lower Energy threshold [keV] | 0.2 | 0.1 | 0.1 |
| Upper Energy threshold [keV] | 2.2 | 2.2 | 3.0 |
| $A_{effective}$[1] @ 0.6 keV [cm²] | 1000 | 1163 | 1400 |
| $A_{effective}$ @ 1.5 keV [cm²] | 400 | 499 | 600 |
| grasp @ 0.6 keV [cm² deg²] | 400 | 405 | 500 |
| Angular res. (HPD) [arcmin] | 4 | 3.7 | 2.5 |
| count rate full FoV [c/s] | 2000 | 2000 | 2000 |
| Source count rate / pixel [c/s] | 10 | 10 | 50 |
| Peak count rate [c/s] | 10,000 | 10,000 | 30,000 |
| Instrumental background @ 1keV [counts cm⁻² s⁻¹ keV⁻¹] | $1.5 \cdot 10^{-2}$ | $5 \cdot 10^{-3}$ | $2 \cdot 10^{-3}$ |

[1] effective area includes mirrors, detector efficiency and filter transmission

## 4.2 Wide-Field Imager (WFI)

To detect point sources at a flux limit of $1.5 \times 10^{-16}$ erg cm$^{-2}$s$^{-1}$ (0.5 – 2 keV), so that we can measure accurately the outer regions of clusters and study the evolution of groups and clusters of galaxies from the formation epoch to the local Universe, the WFI must have a low background and good angular resolution. The low particle background of the MOS CCD technology is proven on XMM, and will be further reduced due to the Low Earth Orbit of EDGE. The cosmic X-ray background comprises a diffuse galactic component and an extragalactic component mostly due to unresolved sources. The angular resolution of the WFImager (HPD < 15") will enable us to resolve the bulk of the latter component. The truly diffuse background and unresolved fraction of the extragalactic component will be characterized by long off-source observations and combining the high angular and high spectral resolution data.

Critical for the WFI is good angular resolution of the mirror (<15") and low scattering wings over the full field of view. These have been achieved using a polynomial approximation of the mirror shapes (5th order polynomial) and a radially tilted focal plane (6°). The use of 2 mm thick SiC mirror shells has been successfully demonstrated at different X-ray beam facilities. We expect that a reduction to 1 mm thickness is a reasonable goal for a mission to be selected in 2011. In this case, while maintaining the good image quality, we will increase the effective area by ~30% at low energies and ~50% at high energies. In Fig. 4.6 we show the expected angular resolution including assembly errors.

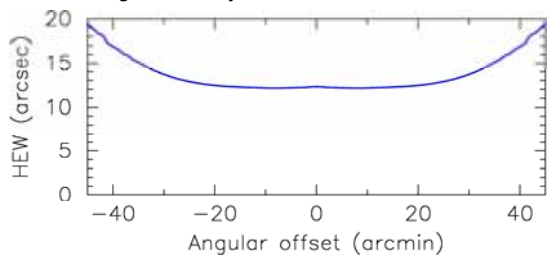

Fig. 4.6 Mirror PSF (HEW) as function of off-axis angle, including assembly errors.

To maintain the angular resolution over the full diameter of the field-of-view (1.5°) a mosaic of partially trapezoidal CCDs tilted by about 6° is used. Thin CCDs, similar to the devices used in *XMM-Newton*, can easily be made in the required format and placed at tilted positions. The proposed tapered image section is of low risk and has also been used on *XMM-Newton* (outside the field of view). The CCDs are passively cooled to -50° C, and a Thermal Electrical Cooler is used to further cool them to the operational temperature of -80° C. The telemetry is dominated by observations of bright Galactic sources and will handle sources up to 0.5 Crab (50 bits/event). Following a GRB, the read-out of the 6 outer CCDs will be suspended, and the inner CCD will switch into a high count rate *imaging mode* (integration times of 0.1 and 2.5 s) or into a *windowed timing mode* (1-D imaging) capable of counting 20,000 counts s$^{-1}$ with acceptable pileup. Readout in these modes is followed by the standard 7-CCD photon counting mode as the GRB intensity decays. A filter wheel, operated at room temperature, will be incorporated to provide filter thicknesses optimized for different science programs. The filter wheel includes a calibration source for in-flight instrument calibration. Contamination control measures similar to those adopted by XMM-EPIC keep contamination of the cooled detectors at acceptable levels.

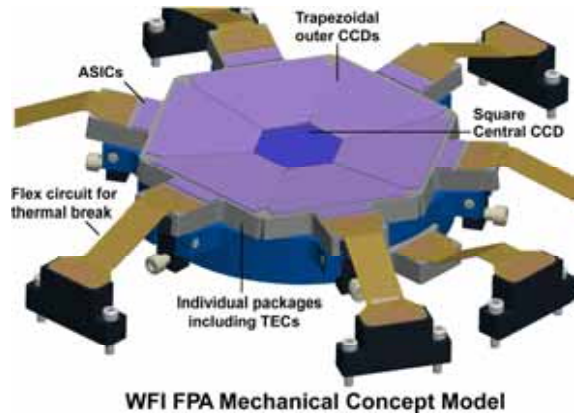

Fig. 4.7 CCD focal plane array design. The inner CCD is below the outer ring and is a standard rectangular image array, whilst the outer ring of CCDs uses the novel trapezoidal image section

The predicted performance is shown in Fig. 4.8 and 4.9, which show the effective area for a number of off-axis angles and the grasp, respectively (including the thermal blanket and detector response). For the trapezoidal CCDs, the Quantum Efficiency (QE) is likely to vary by 5% from low to high off-axis angle, which is <10% of the mirror vignetting function. Both will require detailed modeling and calibration.

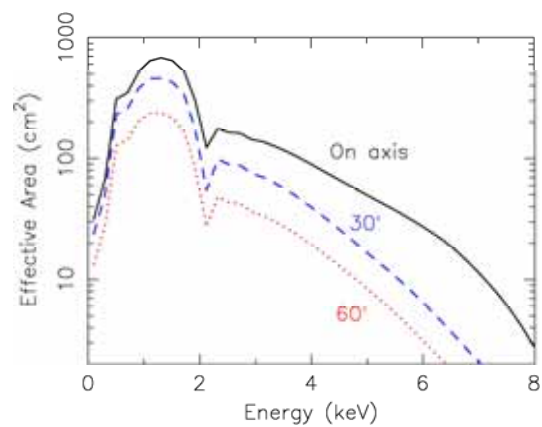

Fig. 4.8 Effective area of the WFI (including mirror, detector QE and thermal blanket) for different off-axis angles

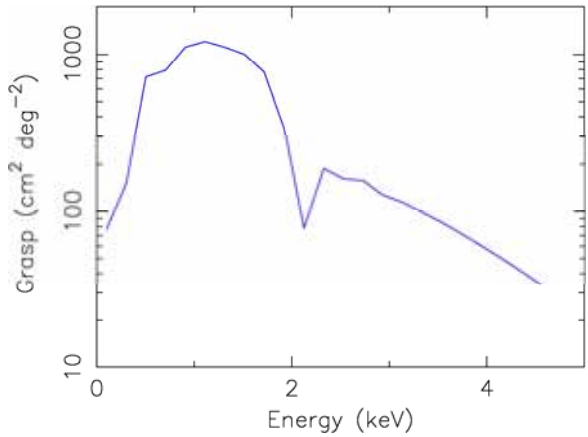

*Fig 4.9 GRASP as a function of energy of the WFI*

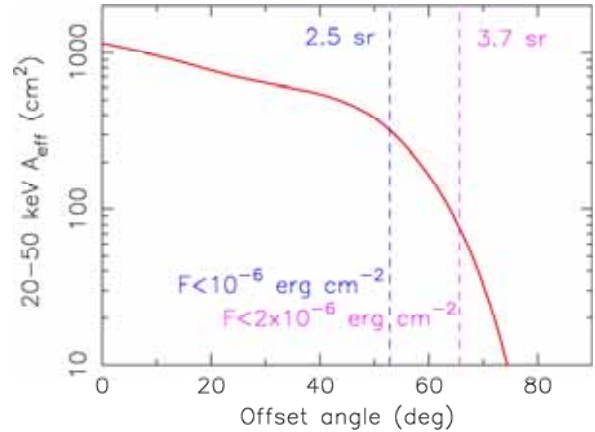

*Fig. 4.10 Effective area of the WFM (2 units) indicating the limiting sensitivity for bursts at different fluences. The fluence limit corresponds to the required sensitivity.*

*Table 4.2 WFI requirements, baseline and goals*

| Parameter | Requirement | Baseline | goal |
|---|---|---|---|
| Resolution[1] at 1 keV [eV] | 80 | 70 | 70 |
| Resolution at 6 keV [eV] | 150 | 130 | 130 |
| FoV (diameter) [deg] | 1.4 | 1.5 | 1.5 |
| Lower threshold [keV] | 0.3 | 0.3 | 0.2 |
| Upper threshold [keV] | 5.0 | 6.0 | 10.0 |
| $A_{effective}$ @ 1 keV [cm²] | 530 | 580 | 1000 |
| $A_{effective}$ @ 3 keV [cm²] | 150 | 140 | 210 |
| $A_{effective}$ @ 6 keV [cm²] | 25 | 25 | 100[2] |
| Angular res. (HPD) [arcsec] | 15 | 15 | 12 |
| source count rate [counts/s] | 1200 | 1200 | 1200 |
| Peak countrate[1] [c/s] | 10.000 | 10.000 | 30.000 |
| Instrumental background @ 1 keV [counts s⁻¹kev⁻¹ cm⁻²] | 1.5 10⁻³ | 1.1 10⁻³ | 6 10⁻⁴ |

[1] The energy resolution is about 250 eV for the windowed timing mode (high count rates)
[2] the goal effective area would require a longer focal length

*Table 4.3 WFM requirements, baseline and goals (2 units)*

| Parameter | requirement | Baseline | goal |
|---|---|---|---|
| ΔE/E @100 keV [FWHM, %] | 5 | 3 | 3 |
| Field of View [sr] | 2.5 | 2.6 | 3 |
| Low threshold [keV] | 8 | 8 | 5 |
| High threshold [keV] | 200 | 200 | 250 |
| On-axis $A_{eff}$ 10-100 keV [cm²] | 1000[1] | 1100 | 1200 |
| Angular res. (HPD) [arcmin] | 35 | 34 | 18 |
| Location accuracy [arcmin] | 4 | 4 | 2 |
| Time resolution [μs] | 10 | 10 | 2 |
| Max rate [counts/s/camera] | 5000 | 5000 | 5000 |
| Software processing time[2] [s] | 20 | 20 | 10 |
| Continuum sensitivity (1s) [ph/cm²/s] | 0.5 | 0.4 | 0.4 |

[1] required averaged area over the FoV is 300 cm²
[2] Depends on S/N

## 4.3 Wide-Field Monitor (WFM)

The Wide-Field Monitor will monitor a ~3 sr solid angle and will localize GRBs with a fluence greater than 10⁻⁶ erg cm⁻² (15-150 keV), with a positional uncertainty < 4'. It will locate a sufficient number of GRB with afterglow bright enough to determine the absorption from the WHIM and to measure the cosmic history of metals at GRB sites. However, lowering the threshold down to 5 keV allows a substantial increase of X-ray flash (XRF) and high-z GRB detections and this increases the number of bright afterglows useful for the WHIM and metallicity studies.

The design is based on two standard coded mask instruments with CdZnTe detectors with good efficiency between 6 and 200 keV (2 mm thick). By tilting these instruments by 28º with respect to the optical axis of the WFI and WFS, the sensitivity of the cameras can be optimized and is consistent with the 2.5 sr coverage, as is shown in Fig. 4.10. For bursts as bright as 2 10⁻⁶ erg cm⁻² the FoV is even 3.7 sr. With a distance to the mask of 40.5 cm and a pixel size of 2.7 mm, a location accuracy of 4' is achieved for sources detected with a S/N > 10. This is sufficient to trigger fast repointing and to locate the source in the 6'x6' field-of-view of the high count rate section of the WFS. The on-board trigger will be based on sampling count rates with different integration times, and includes an imaging trigger based on an on-board catalogue of known sources. This procedure is standard in satellites like SWIFT and AGILE. Following the re-pointing, the data of the WFI can be used to tune the pointing to within 1' (or better). The sensitivity of the camera is sufficient to detect and localize about 80 bursts with a prompt fluence of > $10^{-6}$ erg cm$^{-2}$ per year and is comparable to the IBIS instrument. Similar detectors have been produced (IBIS, BAT) or are currently under development with our partners (Italy, Denmark and France).

### 4.4 Gamma-Ray Burst Detector (GRBD)

The prime goal of the GRB Detector is to measure the high-energy spectrum of GRBs, assessing their potential as standard candles for Cosmology. Furthermore, we expect to use its results to avoid false triggers when we lower the WFM threshold to be more sensitive to high-z GRBs. For the measurement of the high-energy end of the spectrum we need to determine the peak energy $E_p$ with ~10-20% accuracy for about 100-150 GRBs over the mission lifetime. This requires bandwidth extension to around 2 MeV, with sensitivity of about 2 photons cm$^{-2}$ s$^{-1}$ in the 50–300 keV energy band, and co-alignment of this detector with the WFMonitor.

The design includes two scintillators (NaI, 850 cm$^2$ each) read out by 3 PMTs. With a high ratio of photocathode area to scintillator area, the required energy resolution can easily be achieved. With an entrance window of about 0.5 mm thick Aluminum the lower threshold is 20 keV. Both units will be tilted by 45° with respect to each other. Depending on the count rates, onboard spectra are generated with varying integration times (8 s and 16 ms during bursts), or raw events can be stored. The proposed detectors are similar to the GBM on GLAST, which were developed by some of our US partners (MSFC, UAH, USRA), and the technology is widely available.

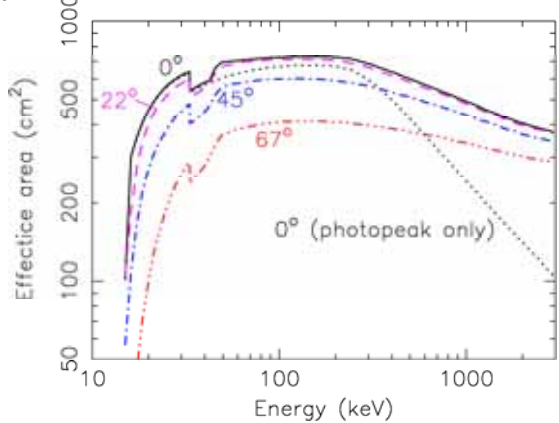

*Fig. 4.10 Effective area of a single GRBDetector.*

Potentially, the GRBDetector and the WFMonitors can be integrated into a single instrument (e.g. a scintillator read out by a Silicon Drift Detector). This would reduce the weight, but this technology is less advanced.

*Table 4.4 GRBD requirements, baseline and goals (2 units)*

| Parameter | Requirement | Baseline | goal |
|---|---|---|---|
| ΔE/E @ 100 keV [FWHM,%] | 20 | 15 | 12 |
| Field of View [sr] | 3 | 4.4 | 4.4 |
| lower threshold [keV] | 25 | 20 | 20 |
| upper threshold [keV] | 2500 | 3000 | 4000 |
| $A_{effective}$ @ 600 keV [cm$^2$] | 800 | 1140 | 1200 |
| Time resolution [ms] | 50 | 8 | 5 |
| Peak count rate per unit [kc/s] | 100 | 200 | 300 |
| Sensitivity in σ (25 s) for standard GRB[1) | | | |
| 100 – 300 keV | 10 | 15 | 30 |
| 300 – 1000 keV | 5 | 5.7 | 15 |
| 1000 – 3000 keV | 1 | 1.1 | 3 |

1)   fluence of $10^{-6}$ erg cm$^{-2}$ in 10-150 keV

### 4.5 Payload generic resources

In this section we describe the alignment requirements, operating modes, calibration approach and the mass/power requirements for the payload. The resources required to accommodate the payload are shown in Table 4.5. Due to the wide field-of-view, the pointing and *alignment requirements* are modest (with the exception of the focal length which is clearly feasible, but needs careful engineering). There are no very demanding requirements on the *operating modes*, aside from the requirement that the WFM/GRB detector combination must be able to trigger the fast repointing using advanced trigger algorithms. The WFI and WFS will run in a photon counting mode whereas the WFM and GRBD will run, depending on the source counts, in a photon counting mode or perform onboard data reduction. To *calibrate* the instruments our ground calibrations will be supplemented with detailed physical models. Ground calibration will be performed by the instrument teams, which have access to different long beam facilities (XACT facility of INAF-OAPA and the long beam facility of MSFC in the US). Modeling, using GEANT, allows for sufficiently accurate calibration of the GRB detector since one unit always has a free field of view relative to the GRB position. To reduce systematic errors due to variations in the different detectors, we foresee active dithering of the satellite resulting in imaging a point of the sky on different detector elements. An in-flight calibration plan that includes revisiting some sources (e.g. SNR, isolated NS) at regular times is foreseen. In addition during the initial phase of the mission, some aspects of the calibration will be verified (e.g. off-axis response). The Technical Readiness Levels (TRL) are all ≥ 4, as substantiated in Section 7. The payload will be provided by institutes in different European countries, as well as by international partners (see Section 8). Although it is recognized that the amount of work required to raise the readiness of the payload from its current level ≥ 4 is significant, we are confident that there are no showstoppers. Some of the enhancements currently under study include:

- the use of thinner and longer mirror shells for the WFI which would increase the area (and reduce cost)
- the use of back illuminated CCDs for the WFI (increasing the QE at low energy at the expense of some extra background)
- a longer focal length of the WFI to enhance its high-energy response
- change of conical approximation for the WFS optics to a Wolter I design improving the angular resolution and effective area
- improvement of the spectral resolution of the WFS, which may be achievable by increasing the number of read-out channels
- Integration of the WFM and GRB detector to reduce the required resources (mass and power) and improve the repointing performance (faster)

*Table 4.5 Payload resources*

| Resource | WFS | WFI | WFM | GRBD |
|---|---|---|---|---|
| # boxes | 7 | 4 | 5 | 5 |
| Bare mass [kg] | 324 | 204 | 111 | 74 |
| Mass+margin [kg] | 389 | 245 | 133 | 89 |
| Bare power [W] | 620 | 140 | 175 | 57 |
| Max power[W] | 679 | 160 | 175 | 57 |
| TM rate [Mbps] | 0.9 | 1.0 | 1.4 | 0.4 |
| TRL | ≥ 4 | ≥ 5 | ≥ 6 | ≥ 7 |

# 5 Spacecraft

The mission profile has been given in Section 3 where some of the key requirements, including the fast repointing, large power and significant cooling requirements were introduced. In this section we illustrate that these demanding requirements can be implemented with state of the art (but also on the shelf) equipment. Therefore there are no critical showstoppers.

## 5.1 Satellite design

The following major points drive the preliminary spacecraft design for the EDGE mission:

- The payload characteristics and requirements: size, mass, power

- The fairing and lift capability of the launcher (2300 kg @ 500 km circular, equatorial)
- The ground station characteristics (S-band, one contact per orbit)
- Selection, as far as possible, of off-the-shelf already qualified equipment and proven technology (TRL>9)
- No procurement risk for components (ITAR free)
- The spacecraft concept keeps the mass as low as possible
- Service Module first priority is the fast repointing capability
- Full redundancy has to be provided for service subsystems
- Simple system architecture concept
- Detailed resource budget to verify compliance with the requirements
- Standard ESA margin philosophy is applied (SCI-A-2003.302/AA)

The main subsystems which have been considered in the preliminary design are:

- structure and accommodation of payload
- Thermal Control System
- AOCS (with fast repointing)
- on-board data handling
- Telecommunication and ground link
- Electrical Power System

These subsystems will all be part of the service module (SVM). An industrial study, funded by ASI, was carried out by Thales Alenia Space and the main results are discussed below following presentation of the satellite configuration.

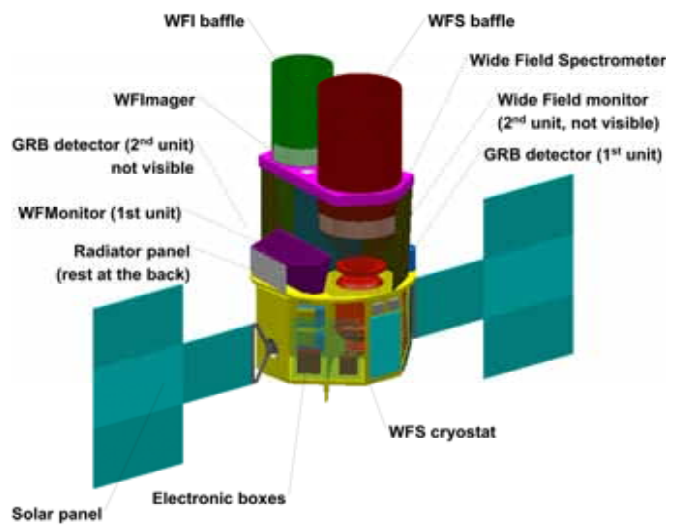

*Fig. 5.1 EDGE payload accommodation. The service module is shown as yellow box. The radiator panels at the bottom and backside of the SVM are not visible.*

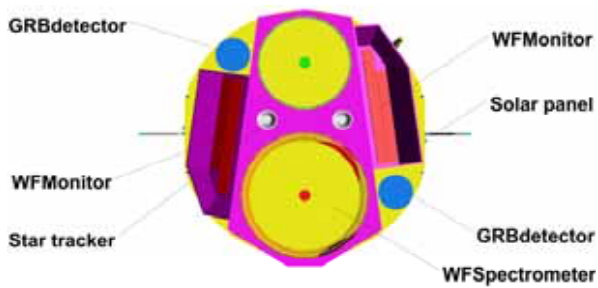

*Fig. 5.2 Top view of the satellite*

## 5.2    Attitude and Orbit Control System

The Attitude and Orbit Control System (AOCS) design is driven by the required Fast Repointing Capability and represents the most mission-specific capability of the SVM. The AOCS has been evaluated assuming two standard slews and one fast slew for each orbit. The main requirements are:

1. Autonomous slewing to a transient event (detected by WFM) with a goal slew rate of 1 deg s$^{-1}$.
2. Acquisition of new transient event locations with an accuracy of 2' within a time frame of 60-120 s for a source 60$^o$ off the original pointing.
3. Pointing accuracy (absolute pointing drift) on longer time scales (>5 min)  in the range of 0.75' to 1.25'
4. Post-facto pointing reconstruction: <3".

*Requirements 1 and 2* resulted in a preliminary trade-off analysis in order to understand which solution is most mature on the basis of the following criteria: compliance with the requirements, mass, power, TRL level, complexity of the control law and command resolution (S/C jitter). *Requirement 3* induces requirements on the platform (thermal stability, selected AOCS equipment), operations (in-flight calibrations) and ground-processing (post-processing algorithms). *Requirement 4* drives the thermal control capabilities, AOCS equipment selection, operations (in-flight calibrations) and ground processing (post-processing algorithms). The trade off for the fast slew actuator selection has been performed considering the following possible solutions:

• Reaction Wheels  (RW) only,
• Reaction Wheels +Thrusters,
• Control Moment Gyro (CMG)

The preliminary evaluation performed by Thales Alenia Space shows that the solution with Reaction Wheels only is the more viable solution. It is technically most mature (compared to CMG) and its weight is acceptable (compared to RW+thrusters). The analysis has been performed for different off-the-shelf reaction wheels and these result typically in a slew time for 60$^o$ of 60-65 s in

80% of the cases, 45–55 s for 50% and a maximum slew time of 85-90 s. The required peak power for this approach (2300 W) can be accommodated by the power subsystem. The post-facto pointing accuracy can be as good as < 0.3-0.5" (1 sigma) using a star tracker with a large FOV (10°x10° or 20°x20°) and one high accuracy gyro which is based on off-the-shelf equipment.

## 5.3    Thermal design

The Thermal Control System (TCS) can be implemented following a standard approach. Considering the low earth orbit and the large power and cooling needs, this is clearly an area which can be significantly optimized by means of a detailed thermal and orbital analysis. The TCS will be developed on two levels: the instrument dedicated thermal control and the service module thermal control.

*Payload thermal control*

A specific analysis, utilizing the Thales Alenia experience, has been performed for the two mirror modules based on 1 node per mirror shell. This specific effort has been necessary because the mirror module dissipations in the S/C are expected to be large because of the large optic apertures. For these apertures we prefer to avoid the use of a thermal filter as this reduces the low energy response of our instruments. Hence we consider a dedicated solution which is different for each telescope.

*Wide-Field Imager*

The CCDs of the WFImager need cooling to -80°C, which is feasible by passive cooling to -50° combined with a Thermo-Electric Cooler. The requirement for the operational temperature of the mirror system is 20°±5°C which translates into a hot external environment to compensate the heat losses to space. This compensation can be obtained by a warmed baffle and the use of a thermal filter in front of the optics (already used on BeppoSAX, Suzaku). A preliminary assessment shows that a 100 cm long baffle with a thermal filter positioned at 20 cm from the mirror module limits the heating power to 40 W. Although this filter reduces the low energy response of the instrument, it reduces the optical load of the detectors as well.

*Wide-Field Spectrometer*

Cooling of the detectors is achieved by a cryogen free cooling system as described in the payload section. No filter on the optics can be included as this would reduce the low energy response unacceptably (the detector already requires 4 thermal filters). Preliminary evaluation shows that the thermal control can be realized with open optics utilizing a baffle height that fits inside the VEGA shroud. An electrical heater network, dissipating the heater power directly on the optics (like XMM), is used to stabilize the temperature with a minimal gradient.

*Service Module Thermal Control Approach*

The TCS of the SVM can be designed on the basis of a standard passive thermal concept. Radiators will be sized to reject the dissipative power in the hot case. The heating power will be sized to maintain the minimum equipment temperature requirement in the cold case. In view of (a) the power dissipation (~1200 W), (b) circular low Earth orbit and (c) internal Service Module environment (-20°/50°C), a radiative area around 4 $m^2$ is needed. This will be localized to the anti-Sun side, lateral panels and lower deck of the SVM. The system can be based on proven solutions like: MLI, radiators, paints, tapes, heaters, thermistors, thermostats and filters, while for the items with high power dissipation the use of heat pipes can be considered. The use of variable conductance heat pipes or heat diodes will be evaluated as thermal switches for thermal disconnection in case of exposure of the radiators to sun or earth.

## 5.4 Avionics: On – Board Data Handling

The S/C functions can be centralized into a single, redundant CDMU (Control and Data Management Unit) unit, which acts as the data handling core controller of the satellite. C&C (Command and Control), AOCS and FDIR (Failure Detection, Isolation and Recovery) functions will be all performed in the CDMU. The assumed CDMU is based on evolution of the Thales Alenia Space Leonardo concept already used in many S/C (RADAR-SAT, COSMO, GOCE, etc.). The Mass Memory is sized at 5 Gbits. AOCS sensors/actuators will have a direct interface via dedicated I/F or via the 1553 data bus. The telemetry/telecommand (TM/TC) unit will have a 1553 data bus, via discrete lines or serial (RS422) links; the link with transponders will be realized via TM/TC dedicated serial links. The instruments will perform their instrument specific data processing including the localization of transient events

## 5.5 Telecommunication

This subsystem includes two main components:

- A standard TM/TC link between S/C and Earth for standard S/C operations. This standard link operates in S-band for a TC rate of about 2-4 kbps. For the TM rates we envisage the use of the high transmission rate (3.8 Mbps), expected to be available for this mission.
- A S/C-ORBCOMM low frequency link for quick look alerts and telecommands. This bidirectional link (already proven for the ASI-AGILE mission) will be used to allow GRB alerts to be rapidly transferred to the ground and to uplink TOO positions.

The data transfer and the scheduled uploading of satellite and payload command loads will be done through the Kourou ground station. The ORBCOMM system is shown in Fig. 5.3.

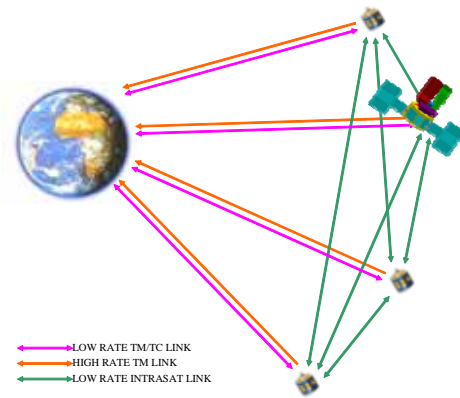

*Fig 5.3    Telecommunication subsystem and architecture of the EDGE-ORBCOMM link*

## 5.6 Electrical power

The Electrical Power Subsystem (EPS) consists of the units/items that will generate/distribute electrical power to the remaining units of the satellite:

- The solar array (total of 11$m^2$, triple junction GaAs) installed on the S/C sunshield, which converts sunlight into electrical power.
- The power conditioning unit, which regulates and controls the solar array power to be compatible with the user characteristics and which provides power distribution control and user protection.
- The battery, which provides power during the Sun eclipses and the launch phases.

The electric power subsystem is sized for 3400 W for the P/L instruments and about 400 W for the SVM. This includes power for charging the batteries and for fast repointing.

The EPS for EDGE is a reliable system that will provide the needed power during sunlight and eclipses, ensuring the best utilization of proven European technology while minimizing mass, volume and used surfaces. Each solar array wing can keep Sun pointing by means of a standard rotator stage. The wings' T-shape represents a solution with standard elements designed to minimize the extension when open, and to stow within the fairing envelope when closed.

## 5.7 Satellite budgets

A detailed breakdown of the payload was made which formed the basis of the accommodation study for the satellite and was also used to estimate mass, volume and power. The top-level hardware tree including all satellite systems has been produced. In Table 5.1 we summarize the performance of the SVM and in Table 5.2 the total satellite budgets are given.

We have identified no critical items but during the next phase detailed modeling of the thermal design and the fast repointing system will have the highest priority.

*Table 5.2  Total satellite mass and power budget*

| parameter | Mass +margin [kg] | Power [W] | Max Power [W] |
|---|---|---|---|
| Payload | 855 | 992 | 1089 |
| Service module | 721 | 353 | 431 |
| System margin[1] | 311 | 134 | 152 |
| Adaptor | 45 | | |
| Total | 1932 | 1479 | 1672 |

1)     20% on mass, 10% on max power

*Table 5.1 Summary of service module subsystems and performance (no margins)*

| Subsystem | Units/key items | budget | Performance |
|---|---|---|---|
| Structure | Primary + secondary structure and adapter (45 kg) | 227 kg | |
| AOCS | 2 Sun sensors, 1 gyroscope, 2 star trackers, 2 magnetometers, 3 magnetic torquers, 4+2 reaction wheels, 2 coarse rate sensors, attitude anomaly detector | 121 kg | • 60° in 60-120 s<br>• <3' pointing accuracy<br>• <1'' AME per axis |
| Thermal | 4 m² radiator Multi-Layer Insulation (MLI), paints, heaters, thermostats, heatpipes, heat diodes(TBC) | 18 kg | • -20° C < $T_{SVM}$ < 50°C<br>• mirror module stabilization<br>• 1200 W power dissipation |
| CDMU | Leonardo concept, 1 Gbyte memory | 20 kg | |
| TT&C | 2 Band X-transponder with travelling wave tube amplifiers<br>via EDGE-ORBCOMM system: TM transmitter, TC receiver, channel diplexer | 13 kg | • up-link 2-4 kbit/s<br>• down-link 3.8 Mbit/s |
| Electrical Power subsystem | Control unit, battery, 11 m² solar panels (3J GaAs) | 206 kg | • 1089 W payload<br>• 431 W SVM<br>• peak power:  2300 W (fast repointing) |
| Total | | 605 kg | |

# 6     Science operations and archiving

## 6.1   Science operations

Science operations will be carried out under the responsibility of ESA. The observation program includes a large core program, from which the data will become public, and a guest observer program (20% of the time during the nominal mission phase) with the usual one year proprietary data rights. About 3% of the time will be reserved for routine in-orbit calibration. The core program will be agreed by a 'science working team' (representatives of the instrument teams and of the science community). It is expected that a major review of the core program will be carried out one year before launch and after the first 1.5 year in operation. In view of the limited GO program during the nominal mission duration (7 month GO time in the first 3 years), two Announcements of Opportunity (AOs) for GO programs will be issued and reviewed by the Observatory Time Allocation Committee.

The scheduling of observations will take advantage of the capability to autonomously repoint the satellite. A list of preplanned targets with good observing efficiency will be uploaded. This target list allows execution of the observing program without ground intervention for up to 3 days. Trigger criteria for the fast pointing can be uploaded from ground. After a valid on-board trigger giving the position of an AT (Automated Target), a fast autonomous slew to the AT will be carried out. The on-board s/w will check that no observing constraints are violated. By default any new AT will be allocated a predefined observing time (50 ks), that can be changed from the ground. Therefore the Science Operations Centre will be based on 09.00 – 17.00 hours, 5 days / week and on-call for weekends.

The instrument/science teams will remain responsible for the health and calibration of their instruments, for defining the trigger criteria and for the fast analysis of GRBs and transients, hence reducing the load for the science operations group. Data from valid GRB triggers and consequent follow/up observations (positions, spectra, light curves) will be transmitted to ground in real time and distributed via the internet to the worldwide community. These data will be inspected on a daily basis by the instrument teams (24 hours, 7 days/week on call).

### 6.2 Data center and archive

For the data center and science archive a similar approach is proposed to that of the INTEGRAL mission. Main tasks of the center include health checking of the science data, routine pipeline processing of the raw instrument data into physical (SI) units (level I including source and background spectra) and archiving of the raw and processed data. Mirror systems will be set up at collaborating international partners (Japan, USA) and in interested European countries. The instrument teams will provide the detailed knowledge and software to process the data according to an agreed standard. Whereas the instrument/science teams have the objective to achieve the prime mission goals, cooperation within the science community at large will be stimulated e.g. by organizing scientific meetings (typically one per year is foreseen on the main science topics). Catalogues based on survey data will be released regularly by the team. The relevant software to reduce the data will also be made available (using heritage of the XMM-Newton mission).

### 6.3 Proprietary data rights

About 80% of the observing time over the first three years will be devoted to the core program. In addition a guest observer program of 20% of the time is planned. The core program data will be made available at regular intervals (once every month) with the exception of the first 6 months. After the commissioning of the satellite, the core program will be initiated. During the first 6 months these data will be analyzed by the instrument teams to sort out the in-flight performance and calibration. Positions of GRBs will be made available through the GRB Coordinates Network once reliable positions and spectra are available. For the guest observer data one year proprietary data rights is guaranteed. Following the nominal duration of this mission we anticipate an extended phase with a more extensive guest observer program.

## 7 Key technology areas

Major advances in space science require either scaling up existing technology (to increase the sensitivity) or applying new, emerging technologies (or a combination of both). For the EDGE proposal we combine proven technology with the recent developments in cryogenic spectrometers. For missions to be launched in 2017, the current TRL level should be at least ≥ 4. This level has already been achieved for the payload and dedicated programs are in progress to raise this to level 6 by the time of the final selection:

- a European/Japanese consortium led by SRON including other partners in the proposal (Italy, Switzerland, Japan, UK and Spain) will realize a full prototype of a TES X-ray micro-calorimeter array in 2008. This prototype is called EURECA.
- For the Japanese mission NeXT (launch 2013) a cooler with similar characteristics (operational temperature of 50 mK) is under development by our Japanese partners.

*Table 7.1 TRL levels of relevant subsystems*

| Unit | Assembly | TRL | Heritage (missions) |
|---|---|---|---|
| **WFImager** | | | |
| | Mirror | 5 | Tests on mirror shells at Panter/MSFC |
| | Detector | 7 | XMM-Newton, Chandra, Suzaku |
| | Filter wheel | 7 | XMM-Newton, many others |
| **WFSpectrometer** | | | |
| | Mirror | 5 | Suzaku, ASCA but some improvements are desirable |
| | Detector | 4 | EURECA tests at Bessy |
| | Read-out | 4 | EURECA (Europe/Japanese prototype) |
| | Cooling | 4 - 5 | Will be operational on NeXT (2013), most sub-assemblies higher |
| | Filters | 5 | Up scaling of filters of Suzaku mission |
| | Filter wheel | 7 | XMM-Newton, many others |
| **WFMonitor** | | | |
| | Detector | 6 | INTEGRAL; ECLAIR (Fr), ASIM (Denmark) fly before EDGE |
| **GRB detector** | | | |
| | Detector | 7 | GLAST, many others |

- Development of the WFI mirrors has been a long program within ASI and INAF. Following the selection of EDGE as one of the three M-class missions, we plan to produce a prototype mirror with a number of shells and to space qualify the performance.

- A program to explore the feasibility of replacing the conical optics for the WFS by a Wolter-I type of optics with better angular resolution and larger area is ongoing.

# 8 Programmatic

## 8.1 Mission management

The proposed mission is a classical ESA science mission for which ESA provides the overall management, the satellite, launch, and science & mission operations, and the national agencies will provide the payload and data processing through an AO and competitive selection. This is illustrated in Fig. 8.1 and Table 8.1. The two X-ray telescopes are split into two separate systems (mirror and detector/cooler) to ensure manageable contributions at the instrument level which can realistically be committed by PI institutes. It is also indicated that the instrument teams maintain an important role during the operational phase (instrument operations, calibration + data processing pipelines, fast analysis of GRBs and repointing priorities).

International collaborations (Japan and US) are expected for substantial parts of the instruments. As a result mirror sites of the processed data will be available in Japan and the USA. In addition, the proposing team includes experts from the *SWIFT* mission, ensuring a good transfer of knowledge from this mission to EDGE. As shown by the accompanying letters, both partners are committed to a pre-phase A study. In addition a number of European space agencies have already explicitly expressed their interest in this mission. In other countries there is a clear interest but the policy was not to provide support letters at this stage of the CV process.

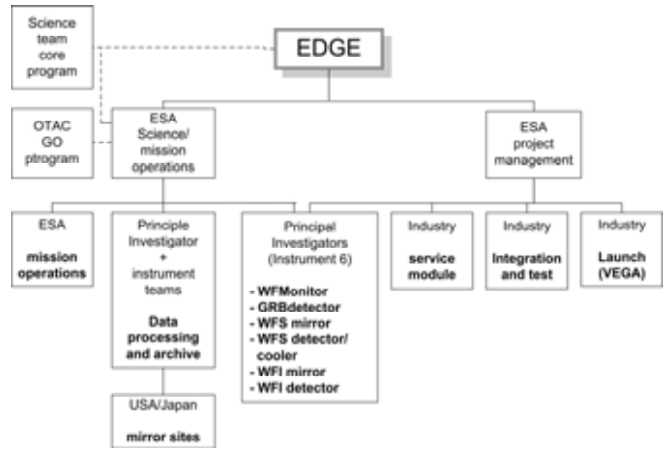

Fig 8.1 Mission management structure. The core observing program is decided by a science team whereas the GO program is selected through open competition. The instrument teams provide the hardware and post-launch support.

.

*Table 8.1 National funded mission components*

| Instrument | Assembly | range [M€] | Cost basis | Cost driver | Interested PI |
|---|---|---|---|---|---|
| WFSpectrometer | Mirror | 10 - 20 | grassroots | Goal of 2.5 arcmin | Japan, Denmark, Italy |
| | Detector | 20 – 40 | grassroots | Goal for > 0.7° FoV | Netherlands, Italy, Japan, USA |
| | cooler | 25 – 40 | grassroots | Redundancy | Japan, UK |
| WFImager | mirror | 21 - 25 | grassroots | Goal of increased high energy area | Italy |
| | Detector | 20 - 24 | Analogy with EPIC-MOS, grassroots | - | UK, USA |
| WFMonitor | 2 units | 40 – 50 | Analogy with other instruments | Goal of 5 keV low level threshold | France,Italy, Denmark |
| GRB detector | 2 units | 15 – 19 | Analogy GLAST, grassroots | - | USA, Japan, |
| Data center[1] | 1 center | 16 - 23 | Analogy INTEGRAL | Operational life | Switzerland, Italy |
| operational support[1] | 6 PI teams | 13 - 18 | Analogy INTEGRAL, AGILE | Operational life | PI teams |
| total | | 180-259 | | | |

1) operational cost range is based on INTEGRAL and is given for a 3 year (low limit) and for 5 year (upper limit) operational lifetime

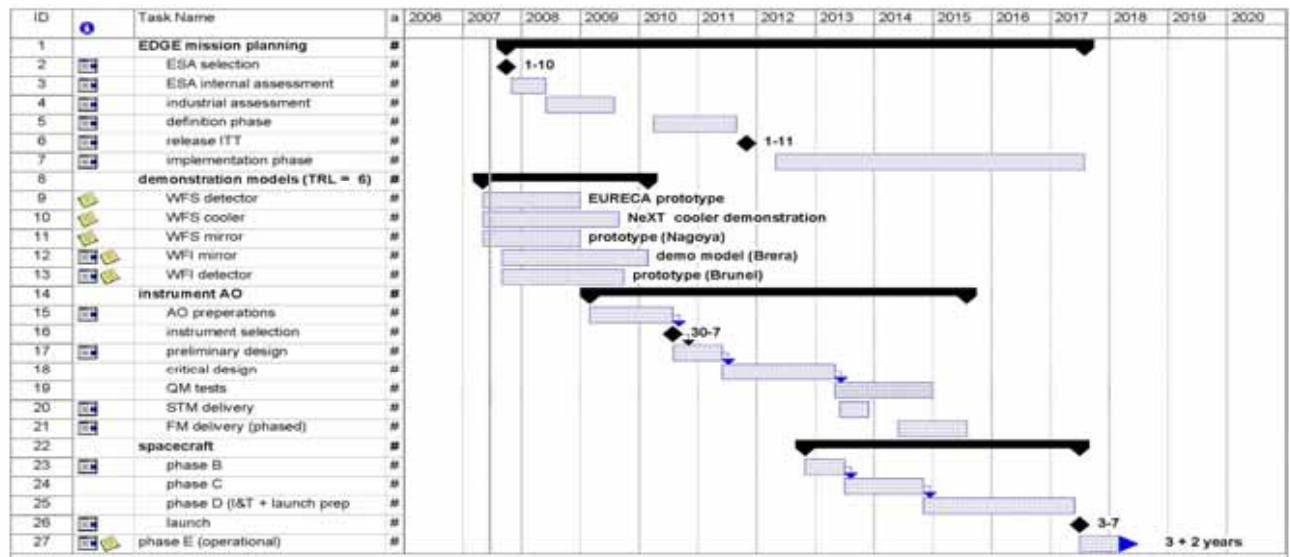

Fig 8.2.   Top level planning

## 8.2   Payload cost

Parametric estimates for the cost of the payload are not very accurate in view of their specialized nature. The cost is estimated based on analogy with existing instruments. Where this is impossible (wide-field spectrometer, mirror of the WFImager) grassroots estimates have been made using a detailed breakdown of the instrument work packages. In these cases we have used a larger range to cope with the related uncertainty.

We are confident that the instrumental contributions will be realized if this mission is selected. For most instruments several institutes are interested in taking the responsibility (Table 8.1) but most likely collaborations between institutes will be formed in view of the magnitude of the required budgets

## 8.3   Mission schedule

The overall mission schedule is shown in Fig. 8.2. Development of critical technology is fully consistent with a selection of EDGE in 2011. Current technology development programs are being carried out for the WFS cooler (development of a cooler for the NeXT mission), for the WFS detector (the EURECA project), the WFS mirror (development at Nagoya University). Development of the technology for the polynomial mirrors is less critical but will be initiated as soon as EDGE is selected as potential M-class mission. The TRL levels will all be > 6 at the time of the final selection.

The key area which will require full study in the early phase includes the Autonomous Spacecraft Control in relation to fast repointing. For this reason, and in order to provide full verification of the AOCS performance, a Fast Repointing Simulator will be developed, at software level only, providing full simulation of the AOCS–Fast Repointing performance.

The TRL levels of the various sub-systems are listed in Table 7.1, including the relevant heritage. The TRL levels are given for the baseline payload. To achieve the desired performance (goals) further engineering of the current technology is required. It should also be noted that some of the technology is more advanced in some non-European countries but in these cases interest to participate/contribute has been expressed (Japan, USA).

## 8.4   Mission cost

Mission costs are estimated in Table 8.2. For a significant fraction of the mission cost we have adopted the cost recommended by ESA. The cost of the Service Module has been estimated following a parametric model (see Space Mission Analysis & Design, Larson and Wertz) and is split into phase A/B/C and phase D/launch preparations. An independent verification by the Engineering Cost Group of Marshall Space Flight Center gives a significantly lower number (105 M$ instead of 138 M€). Comparison with the costs of the *SWIFT* mission suggests that significant reductions of the cost might be feasible (total *SWIFT* budget 180 M$, operated by a operations team of ~ 35 fte/year on a 8.00 – 17.00 basis making full use of the satellite autonomy). Clearly the variance in these numbers indicate that a detailed study is needed but we have adopted the most conservative estimates for our proposal.

*Table 8.2 Mission cost to ESA*

| Activity | Cost range [M€] | Cost basis | Cost driver |
|---|---|---|---|
| Study phase | 6 | 2% (CV spec) | |
| ESA management | 31-33 | 11% (CV spec) | |
| Phase A/B/C | 46 | Parametric model | |
| Phase D + launch prep. | 92 | Parametric model | Cryogenics, cleanliness |
| Launch | 24 | VEGA launch | |
| Mission operations | 17-29 | 3+2 years, using INTEGRAL(5.8/yr) | Operational life |
| Science operations | 7-11 | 3+2 years, using INTEGRAL(2.2/yr) | Operational life |
| contingency | 50-54 | 18% (CV spec) | |
| Public outreach | 1-2 | 0% (CV spec) | |
| Total | 274-297 | | |

## 9 Communication and outreach

The EDGE mission has a clear public appeal that derives from the investigation of questions that are at the same time fundamental for astrophysics and cosmology and that also tantalize the imaginations of lay people. Where are the missing baryons? Was a Gamma-Ray Burst the explosive death of the first stars formed in our Universe? What are the extreme states of matter? How are metals formed and distributed in the Universe? We are committed to an active Education and Public Outreach (EPO) program. Our EPO plans focus on high-leverage programs involving teachers, students, amateur astronomers and the public.
.
We plan to hold teacher workshops in the ESA member states emphasizing the high-energy universe, using the master-teacher approach; to exploit the multiplier effect we will provide materials for the participants to help other teachers in their own school districts. We will hire graduate students in education to review and evaluate these workshops and make suggestions for future meetings.

We will form partnerships with amateur astronomy groups and will hold 2 - 3 day biannual astronomy camps and/or workshops specifically for amateur astronomers. It is worth recalling that amateur astronomers are actively involved in optical follow up observations of GRBs. This type of workshop/astronomy camp and successful profes-sional-amateur collaboration specifically designed to inform and educate the amateur astronomer community was pioneered by the MSFC members of the EDGE collaboration, who have held 3 very successful workshops so far.

We will disseminate our results to the local public by creating planetarium shows and public talks by the EDGE scientists that highlight X- and gamma-ray astronomy. Naturally partnerships with the public outreach office of ESA and the respective national space agencies, which are already providing basic science materials to these areas, are important and we will work together to improve these programs.

We are excited about our proposed public outreach program, for it addresses wide and diverse audiences and has a strong multiplier effect. We expect to concentrate the activities starting one year before launch and slowly fading out after the nominal mission lifetime. This enables a more concentrated effort with relatively larger impact. We have budgeted sufficient resources within EDGE to meet these public outreach goals and we will make full use of the expertise in our collaboration (especially with our US colleagues).

## 10 Cosmic Vision compliance

In this section the EDGE proposal is compared with the criteria as presented by ESA during the CV briefing:

- *Scientific excellence*: EDGE will for the first time measure the low density part of large scale structures in the Universe and investigate their formation and physics. It will probe star formation and explosive metal enrichment by GRBs throughout the Universe. These items have been ranked highly in the CV program and more recently in ASTRONET.

- *Scientific return*: by a combination of a large core program with deep observations of a part of the sky and with GRB follow-up measurements and a modest Guest Observer program, the scientific return will be large. The data of the core program will be made available to the scientific community directly.

- *CV priorities*: this mission will address major parts of the sub-questions 4.2 (the Universe taking shape), 3.3 (matter under extreme conditions) and can, potentially contribute to 2.1 (solar system) and 3.2 (gravitational wave Universe)

- *Time line* (feasibility): Most technologies have a well proven flight heritage with the exception of the cryogenic detector. A similar detector is part of the NeXT mission (launch 2013) although the detector read-out will be different. The smart combination of the proposed payload on a fast repointing satellite (already proven by the *SWIFT* mission) is challenging but certainly feasible within the available time frame

- *Requiring space*: soft X-rays are absorbed in the atmosphere and require satellites or rockets

- *Science for money*: the proposed mission is a medium class mission which will serve a large fraction of the high-energy community. The instruments will be provided by a large and experienced community.

- *Technological maturity*: the current TRL levels are ≥ 4 and for most parts significantly higher. Raising the TRL level of the key technology (detector read-out and coolers and optics) is already funded at the various institutes and is expected to mature in the next 1-2 years.

- *Compatibility with the M envelope*: Mass and budgets are estimated to be within the M envelope taking into account ESA management, launch and operations cost. Comparison with the costs of *SWIFT* (USA/Italy/UK) confirms this conclusion.

- *Cost to the member states*: significant cost to the member states is expected but they already invested heavily in the technologies and the proposed instruments have been defined by groups from different institutes. They will be eager to provide the hardware.

- *Project risks*: the requirements can be met with realistic performance of the instruments. Improved performance is likely by further optimization (but this is not required). Considerable contingencies have been identified (e.g. contingency, cost ranges) and some descoping is feasible without affecting the core science. This is, however, not required.

- *International cooperation*: there is significant interest in Japan and the USA and they contributed as equal partners to the proposal. If selected, major hardware contributions from these partners will be proposed.

- *Communication potential*: the fundamental questions addressed by this mission will be of great interest to the general public. We plan, in addition to national programs, a special program to support this. We expect to make full use of the experience of our colleagues in the USA.

## Abbreviations

| | |
|---|---|
| AGN | Active Galactic Nucleus |
| AO | Announcement of Opportunity |
| AOCS | Attitude and Orbit Control System |
| CDMU | Central Data Management Unit |
| CMG | Control Moment Gyro |
| CV | Cosmic Vision |
| DE | Dark Energy |
| DM | Dark Matter |
| EOS | Equation of State |
| EM | Electro Magnetic |
| EPO | Education and Public Outreach |
| EURECA | European/Japanese consortium building a TES calorimeter prototype |
| EW | equivalent width (area of the line wrt the continuum) |
| FoV | Field of View |
| GO | Guest Observer |
| GRASP | product of effective area and field of view |
| GRB | Gamma-Ray Burst |
| GRBD | GRB Detector |
| GW | Gravitational Wave |
| HPD | Half Power Diameter |
| ICM | Intra Cluster Medium |
| ISM | Inter Stellar Medium |
| LEO | Low Earth Orbit |
| NS | Neutron Star |
| PMT | Photo Multiplier Tube |
| QE | Quantum Efficiency |
| SAA | South Atlantic Anomaly |
| S/C | Spacecraft |
| SFXT | Supergiant Fast X-ray Transient |
| SMBH | Super Massive Black Hole |
| SNR | SuperNova Remnant |
| SVM | Service Module |
| SWCX | Solar Wind Charge eXchange |
| TCS | Thermal Control System |
| TDRS | Tracking and Data Relay Satellite (USA) |
| TES | Transition Edge Sensor |
| TM/TC | Telemetry / Telecommand |
| TOO | target of opportunity |
| TRL | Technical Readiness Level |
| TT&C | Telemetry, tracing and Control |
| WFI | Wide-Field Imager |
| WFM | Wide-Field Monitor |
| WFS | Wide-Field Spectrometer |
| WHIM | Warm-Hot Intergalactic Medium |
| XRF | X-ray Flashes |

## References

The proposal is based on detailed reports on the science, on the instruments and on the satellite. These are available on the EDGE web site (http://projects.iasf-roma.inaf.it /edge/).